# Special scattering regimes for conical all-dielectric nanoparticles

Alexey V. Kuznetsov[1,3,*], Adrià Canós Valero[1], Hadi K. Shamkhi[1], Pavel Terekhov[4], Xingjie Ni[4], Vjaceslavs Bobrovs[3], Mikhail Rybin[1], Alexander S. Shalin[2,3*]

[1]ITMO University, Faculty of Physics, St. Petersburg, 197101, Russia

[2]MSU, Faculty of Physics, Moscow, 119991, Russian Federation

[3]Riga Technical University, Institute of Telecommunications, Riga, 1048, Latvia

[4]The Pennsylvania State University, Department of Electrical Engineering, Pennsylvania, 16802, United States

*alexey.kuznetsov98@gmail.com

*alexandesh@gmail.com

## Abstract

All-dielectric nanophotonics opens a venue for a variety of novel phenomena and scattering regimes driven by unique optical effects in semiconductor and dielectric nanoresonators. Their peculiar optical signatures enabled by simultaneous electric and magnetic responses in the visible range pave a way for a plenty of new applications in nano-optics, biology, sensing, etc. In this work, we investigate fabrication-friendly truncated cone resonators and achieve several important scattering regimes due to the inherent property of cones - broken symmetry along the main axis without involving complex geometries or structured beams. We show this symmetry breaking to deliver various kinds of Kerker effects (Generalized and Transverse Kerker effects), non-scattering hybrid anapole regime (simultaneous anapole conditions for all the multipoles in a particle leading to the nearly full scattering suppression) and, vice versa, superscattering regime. Being governed by the same straightforward geometrical paradigm, discussed effects could greatly simplify the manufacturing process of photonic devices with different functionalities. Moreover, the additional degrees of freedom driven by the conicity open new horizons to tailor light-matter interactions at the nanoscale.

## Introduction

Recently, ever-increasing attention has been paid to the optical properties of subwavelength dielectric and semiconductor nanoparticles. Such objects can be used to create various structures and platforms either significantly outperforming existing ones, or even belonging to a completely new class. Their undeniable advantages are low losses, high efficiency, and the ability to simultaneously tailor both the electric and magnetic components of light. Nanolasers[1], nanoantennas[2], ultrathin lenses[3], sensors and detectors[4–7], metamaterials and metasurfaces[8–10] and others emerging applications of the novel peculiar effects[11–16] are driven by such resonators.

A big step in high-index dielectric nanophotonics was made by implementing a flexible control over multipole excitations in subwavelength scatterers, understanding and tuning different phenomena by the means of multipole moments combinations and interference[17,18]. For example, it became possible to implement the so-called *Kerker effect* – the cancellation of backward or forward scattering from a nanoparticle, which was originally introduced for a hypothetical sphere with equal epsilon and mu[19]. Nowadays, the Kerker effect plays a crucial role in dielectric nanophotonics, giving rise to a variety of fully transparent phase-tailoring "Huygens" metasurfaces[20–22]. As a next step, the so-called generalized Kerker effect was introduced; this effect allows one to observe an interference picture between resonantly excited electromagnetic multipoles of different orders[23–26]. Expanding the possibilities of the usual Kerker condition, it allows governing the scattering directivity via the interplay of multipolar channels. In the recent paper[27] the transverse Kerker effect was shown and verified experimentally; this effect is characterized by isotropic transverse scattering with simultaneous nearly full suppression of both forward and backward scattering. This peculiar optical signature of specially designed nanoparticles enabled fully transparent metasurfaces or even a perfect absorber[28,29].

Another intriguing possibility is an access to the so-called *anapole* (cancellation of the dipole radiation via the toroidal dipole one)[30] or *hybrid anapole* regimes (HA)[31] – simultaneous fulfillment of anapole conditions for all the main multipoles in a scatterer, giving rise to the far-field scattering suppression and strong near-field energy localization regardless of a substrate. Metasurfaces based on such meta-atoms feature many useful properties such as a controlling phase shift of transmitted light, almost perfect transparency, strong nearfield enhancement useful for nonlinear and Raman applications, complex and unusual transient properties, etc., for various types of substrates and shapes[32–34].

Besides the completely non-scattering nanoobjects, researchers are also interested in the cases, when nanoparticles become anomalously strong scatterers. Such states are the opposite of an anapole, because at a certain frequency scattering maxima of several multipoles appear; this state is called *superscattering*[35]. Nanoscatterers supporting this feature can find their application in a number of areas, such as sensing, optical communication, and other emerging areas[36–38].

The cases above are usually considered for highly symmetric particles with simple geometry like spheres, cubes, or cylinders, each of them supporting a limited set of multipole interactions depending on size, material, and aspect ratio. Often, to get specific multipole configuration it is necessary to require some complications, for example multilayer structure. Breaking the symmetry enables more careful mode engineering allowing, for example, truncated conical scatterers (Figure 1) to support all the aforementioned responses with unprecedented flexibility in scattering pattern tailoring without the use of complicated material configurations. Moreover, such geometry has many advantages in terms of fabrication, since most of the conventionally fabricated cylinders are actually cones with small side slope[39–41]. To the date, there is a plenty of studies on conical particles from resonant reflectors[42], color filters[43], and nanoantennas[40] to antireflecting and light-trapping coatings for photovoltaic cells[44], and photonic nanojets[45].

Truncated cones are gaining popularity as elements of photonic structures due to their undeniable advantages. The main such advantage is the opportunity to vary the upper radius, adding an additional degree of freedom. Unlike the more popular cylinder-shaped nano-scatterers, nanocones allow you to vary the mode composition of the scatterer more precisely, which leads to a better tuning and obtaining effects that were previously unattainable[46]. For example, in[42], the authors were able to shift the reflection maximum of the metasurface by varying the geometric dimensions of the truncated nanocone meta-atom, which allowed them to move the resonance of interest through the entire visible spectrum. Additionally, in[43], the authors showed absorptive-type filters using truncated-cone hyperbolic metamaterial absorbers, where the desired operation frequency was achieved by varying the geometry and composition of the meta-atoms.

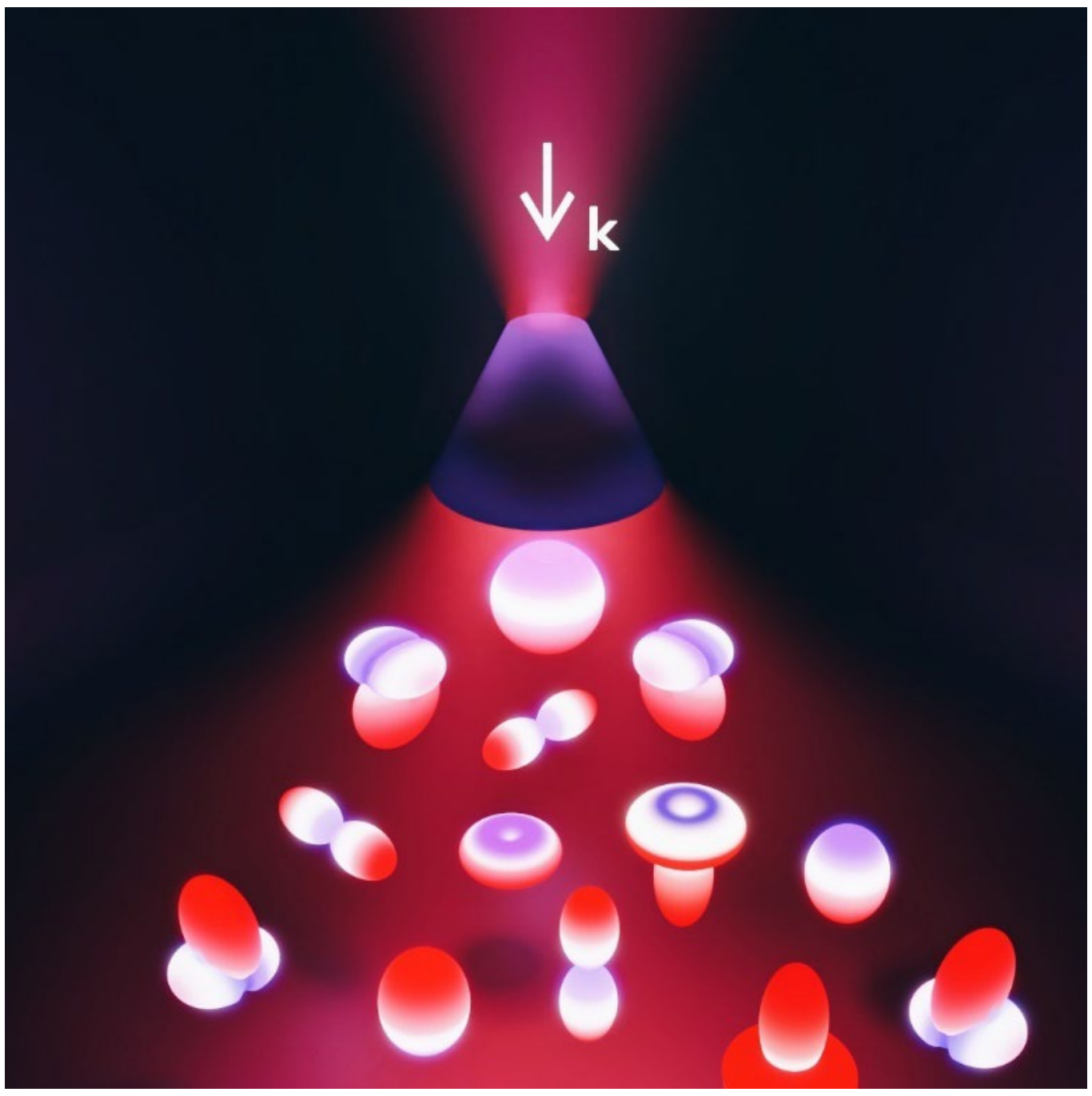

**Figure 1.** Artistic representation of the considered silicon nanocone particles illuminated with the linearly polarized plane wave, showing the main advantage of the truncated cone geometry over the simpler shapes: the possibility of obtaining various important multipole effects within the same geometric shape.

Unfortunately, this geometry is still understudied compared with the "conventional" spherical, cubic, and cylindrical shapes. To bridge this gap, this work presents an extensive tutorial study of the optical properties of truncated conical scatterers. In showcasing of special scattering regimes enabled by the additional degree of freedom - broken symmetry along the main axis. In this work, we demonstrate all the phenomena mentioned above using silicon truncated cone scatterers (experimental dispersion data for a-Si is shown in Supplementary Fig. S1) and indicate the specific parameters corresponding to these effects, for which we performed numerical simulations in Comsol Multiphysics. Based on these data, it is easy to find the desired effect in other materials and at other wavelengths with other geometric values, which makes the approach quite universal.

# 1. Modes and multipoles

## 1.1. Multipolar expansion

Multipole decomposition is an important tool for analyzing the interaction of light with matter. Recently new expressions for multipole moments were introduced, valid for arbitrarily sized particles of any shape. Beyond the particular case of multipolar moments induced by an incident field in a structure, these expressions can be directly applied in the many areas, where the multipole decomposition of electrical current density distributions is used[18]. Exact multipole moments (1)-(4):

$$p_{\alpha} = -\frac{1}{i\omega}\left\{\int d^3\mathbf{r} J_{\alpha}^{\omega} j_0(kr) + \frac{k^2}{2}\int d^3\mathbf{r}\left[3(\mathbf{r}\cdot\mathbf{J}_{\omega})r_{\alpha} - r^2 J_{\alpha}^{\omega}\right]\frac{j_2(kr)}{(kr)^2}\right\}, \tag{1}$$

$$m_\alpha = \frac{3}{2}\int d^3\mathbf{r}(\mathbf{r}\times\mathbf{J}_\omega)_\alpha \frac{j_1(kr)}{kr}, \tag{2}$$

$$\begin{aligned} Q^e_{\alpha\beta} = -\frac{3}{i\omega}\{\int d^3\mathbf{r}\left[3(r_\beta J_\alpha + r_\alpha J_\beta) - 2(\mathbf{r}\cdot\mathbf{J}_\omega)\delta_{\alpha\beta}\right]\frac{j_1(kr)}{kr} + \\ +2k^2\int d^3\mathbf{r}\left[5r_\alpha r_\beta(\mathbf{r}\cdot\mathbf{J}_\omega) - (r_\alpha J_\beta + r_\beta J_\alpha)r^2 - r^2(\mathbf{r}\cdot\mathbf{J}_\omega)\delta_{\alpha\beta}\right]\frac{j_3(kr)}{(kr)^3}\}, \end{aligned} \tag{3}$$

$$Q^m_{\alpha\beta} = 15\int d^3\mathbf{r}\left\{r_\alpha(\mathbf{r}\times\mathbf{J}_\omega)_\beta + r_\beta(\mathbf{r}\times\mathbf{J}_\omega)_\alpha\right\}\frac{j_2(kr)}{(kr)^2}, \tag{4}$$

where $p_\alpha$ – electric dipole moment (ED), $m_\alpha$ – magnetic dipole moment (MD), $Q^e_{\alpha\beta}$ – electric quadrupole moment (EQ) and $Q^m_{\alpha\beta}$ – magnetic quadrupole moment (MQ), $\alpha, \beta = x, y, z$, $\mathbf{J}_\omega$ – current density inside the particle, $j_{1,2,3}$ – spherical Bessel functions of the first kind.

Hereinafter we will not consider any higher order terms [18], since they are negligible in all the cases of interest.

In the far field region, the exact induced multipole moments produce an electric field given by [47]:

$$\mathbf{E} = \frac{k^2}{4\pi\varepsilon_0}\frac{e^{ikr}}{r}\left\{\mathbf{n}\times(\mathbf{p}\times\mathbf{n}) + \frac{1}{c}(\mathbf{m}\times\mathbf{n}) - \frac{ik}{6}\mathbf{n}\times\left[\mathbf{n}\times(\mathbf{Q}^e\cdot\mathbf{n})\right] - \frac{ik}{6c}\left[\mathbf{n}\times(\mathbf{Q}^m\cdot\mathbf{n})\right] + \cdots\right\} \tag{5}$$

Importantly, upon making the change $\mathbf{n}\rightarrow-\mathbf{n}$ in equation (5), the electric field produced by the ED does not change sign, while the electric field produced by the MD is reversed. For quadrupoles, the opposite happens, i.e., the MQ field is even while the EQ one is odd. The 'even-odd' character under space inversion is often referred to as the *parity* of the multipole field. This well-defined behavior under space inversion is at the core of all multipolar interference effects.

We also need to write down the following formula for the scattering cross-section (SCS)[18], to directly compare the contributions of different multipoles to the SCS:

$$\begin{aligned} C^{total}_{sca} = C^p_{sca} + C^m_{sca} + C^{Q^e}_{sca} + C^{Q^m}_{sca} + \cdots = \\ = \frac{k^4}{6\pi\varepsilon_0^2|\mathbf{E}_{inc}|^2}\left[\sum_\alpha\left(|p_\alpha|^2 + \left|\frac{m_\alpha}{c}\right|^2\right) + \frac{1}{120}\sum_{\alpha\beta}\left(|kQ^e_{\alpha\beta}|^2 + \left|\frac{kQ^m_{\alpha\beta}}{c}\right|^2\right) + \cdots\right] \end{aligned} \tag{6}$$

Interestingly, we note that the SCS of each multipole is decoupled from the rest. Thus, every multipole can be viewed as an independent 'scattering channel', through which the nanoparticle can exchange power with the environment[48].

## 1.2. Quasinormal Modes

In the presence of an exciting field, dielectric nanoparticles have the ability to confine light within their volume. In other words, they act as subwavelength nanoresonators, supporting 'resonant modes'. This confinement becomes optimal near their resonant frequencies. To understand the nature of this phenomenon, one can trace an analogy with Fabry-Perot cavities. Once light enters the cavity, it bounces back and forth due to reflection from the walls until it finds a way to exit. However, for some special frequencies, a standing wave can form due to constructive interference, leading to resonant behavior. Another prominent analogy can be found in dielectric microcavities supporting whispering gallery modes, arising by total internal reflection after a roundtrip around the cavity. Resonances in dielectric nanoparticles can be visualized as a mixture between the two examples above; they can arise from standing waves along their transverse dimension (similar to whispering gallery modes), their longitudinal one (resembling Fabry Perot cavities), or a combination of both.

In an ideal system, such as a Fabry Perot resonator bounded by two perfectly conducting mirrors, light can never escape and will oscillate back and forth from the walls for an infinite time. The resonant frequencies in this context are real numbers. However, in an actual physical system, energy eventually dissipates into the environment, either by absorption or radiation. The resonant frequencies are then complex, implying that no stationary excitation can fully access them. Despite this, the system's response (e.g., scattering) is strongly affected when approaching a resonant frequency along the real axis. Thus, understanding their behavior and radiation characteristics is of great importance for the design of nanophotonic devices.

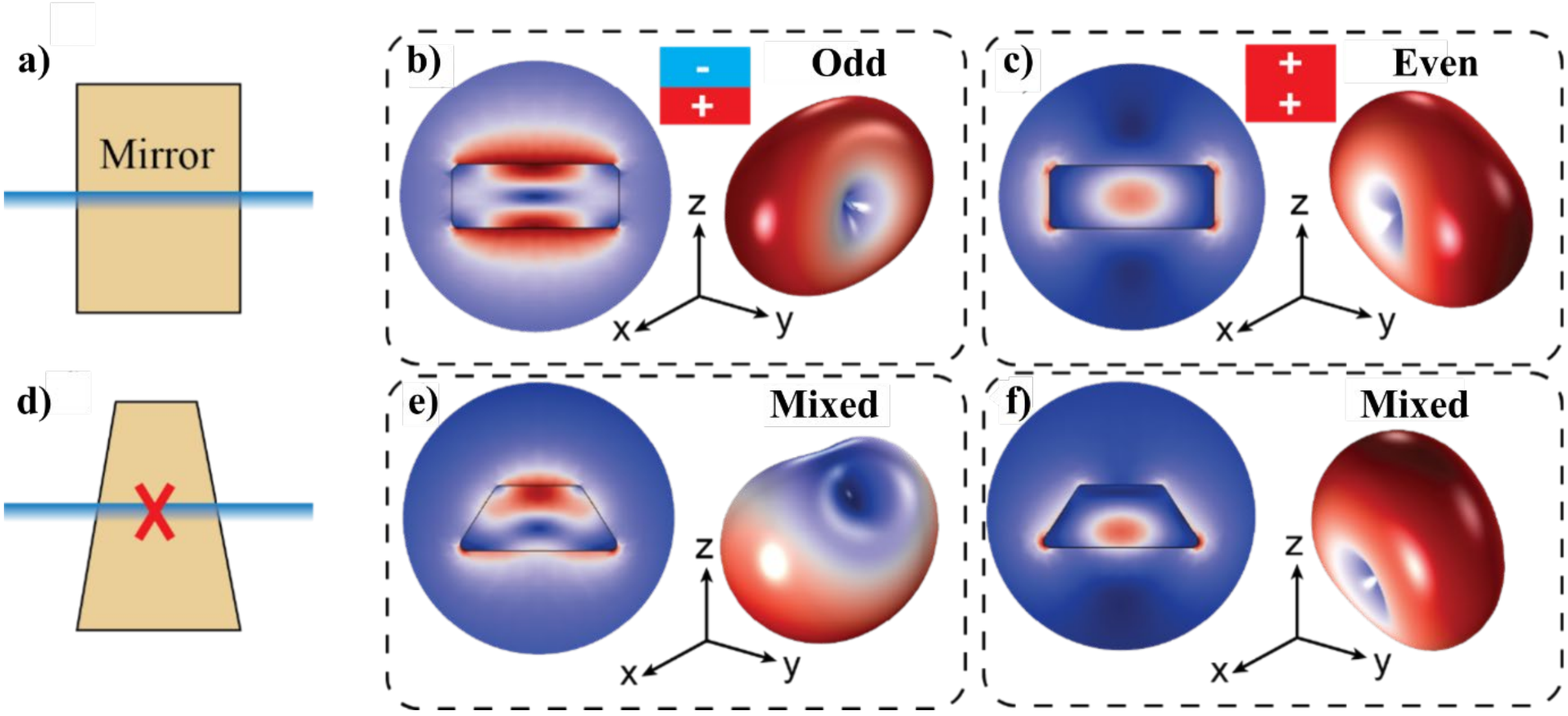

**Figure 2.** Lowest order resonant QNMs of a Si nanocylinder and a truncated Si nanocone. We consider only the QNMs that can be excited by a normally incident, x-polarized plane wave. Parameters of the nanocylinder: height 100 nm, radius 130 nm. Parameters of the nanocone: height 100 nm, bottom radius 130 nm, upper radius 65 nm. The refractive index of both nanoresonators is set to 4. (a) Scheme depicting the x-z cross-section of a cylinder, and it's up-down mirror symmetry. (b) Lowest order QNM of the nanocylinder with odd parity. It can be easily identified as the well-known MD mode. Left: near field distribution in the x-z plane, right: radiation pattern. (c) Lowest order QNM with even parity, corresponding to the ED mode. Left and right pictures as in (b). (d) Representation of the x-z cross-section of a nanocone, indicating the lack of up-down mirror symmetry. As a result, the QNMs are no longer even or odd. In particular, the lowest order QNMs radiate as a combination of ED and MD multipoles. (e)-(f) Lowest order resonant QNMs of the nanocone. Left and right pictures display the same as in (b) and (c). As a result of the mixed electric and magnetic contributions, the forward and backward directivities of a single QNM can be boosted with respect to the nanocylinder.

Formally, resonant modes are eigensolutions of Maxwell's equations, supplemented with a set of boundary conditions. For an isolated dielectric nanoparticle in a homogeneous environment, the boundary conditions can be replaced by Sommerfeld radiation conditions. Assuming a dispersionless permittivity $\varepsilon(\mathbf{r})$, the complete system of equations reads as (we omit for brevity the spatial dependences in the fields and the permittivity):

$$\begin{pmatrix} 0 & -i\nabla\times \\ -i\nabla\times & 0 \end{pmatrix}\begin{pmatrix} \mathbf{E}_m \\ \mathbf{H}_m \end{pmatrix} = \tilde{\omega}_m \begin{pmatrix} \varepsilon & 0 \\ 0 & -\mu_0 \end{pmatrix}\begin{pmatrix} \mathbf{E}_m \\ \mathbf{H}_m \end{pmatrix}, \tag{7}$$

together with the radiation condition:

$$\hat{\mathbf{r}}\times\nabla\times\mathbf{E}_m \to -\mathrm{i}\frac{\tilde{\omega}_m}{\mathrm{c}}\mathbf{E}_m, \quad r\to\infty, \tag{8}$$

where $\hat{\mathbf{r}}$ is a unit radial vector and $\tilde{\omega}_m = \omega_m - i\gamma_m$ is the resonant frequency or *eigenfrequency* associated with the eigenmode $\begin{pmatrix} \mathbf{E}_m \\ \mathbf{H}_m \end{pmatrix}$. Equation (8) essentially tells that the mode field must behave as a spherical wave at infinity, i.e. follow a dependence $\mathbf{E}_m \sim \mathbf{E}_0(\hat{\mathbf{r}})e^{i\tilde{\omega}_m r/c}/r$. Interestingly, to ensure energy decay in the time domain, $\gamma_m$ is always a real, positive number. Therefore, far from the origin, the mode amplitude blows up due to a term $e^{\gamma_m r/c}/r$. In consequence, modes in open systems cannot be normalized by standard means, and often receive the name 'quasinormal' modes (QNMs). Despite this, the scattered field everywhere outside the nanoparticle can be accurately described as a linear combination of the contributions of several QNMs, i.e. $\mathbf{E}_{sca} = \sum_m \alpha_m(\omega)\mathbf{E}_m$, where $\alpha_m(\omega)$ describes the coupling of the m-th QNM to the incident field [49].

Multipolar resonances in dielectric nanoparticles are associated with a QNM. In the case of spheres, every resonant mode radiates as a specific multipole. However, this is no longer true for arbitrary shapes. In a general setting, the scattering pattern of a QNM can be described as a mixture of multipole contributions. To determine how much a QNM 'matches' a given multipole, (i.e. its multipolar content) one can define the m-th 'eigen' current $\mathbf{J}_m = -i\tilde{\omega}_m\varepsilon_0(\varepsilon_p - 1)\mathbf{E}_m$ ,where $\varepsilon_p$ is the relative permittivity of the nanoparticle. Introducing $\mathbf{J}_m$ into equations (1) - (4), and evaluating them at the *complex* frequency $\tilde{\omega}_m$ yields the desired multipolar content of the m-th QNM[50]. At a *real* frequency $\omega$, the QNM will radiate as a combination of its intrinsic multipole moments.

However, the particle symmetry imposes strict bounds on the multipolar contents of its QNMs. For instance, cylindrical symmetry prevents a mixture of multipoles with even and odd parity (refer to Figure 2a-c). This has important implications, e.g., for the design of Kerker meta-atoms: since the electric and magnetic dipoles have opposite symmetry, the radiation pattern by a single mode cannot be directional. Thus, engineering a resonant Kerker effect in a dielectric nanodisk requires overlapping the resonances of two QNMs with electric and magnetic dipolar character, respectively.

Unlike cylinders, truncated nanocones lack the vertical mirror plane, as schematically depicted in Figure 2d. As a result, the QNMs can radiate as a combination of multipoles with even and odd parity, such as the electric and magnetic dipole. Consequently, the directivity of a single QNM can be enhanced in the forward or the backward direction, as can be appreciated from the radiation patterns of the two lowest order QNMs of a nanocone (Figure 2e-f). This also immediately implies the appearance of a bianisotropic response. Thus, truncated nanocones not only hold an additional degree of freedom (conicity) with respect to more studied geometries such as cylinders: conicity provides a simple strategy to control the even-odd mixture of multipoles in a QNM, leading to new exotic effects, such as single mode directivity, strong coupling and exceptional points[51].

## 2. Kerker effects

In this section, we talk about various types of Kerker effects (Conventional, Generalized, Transverse Kerker effects) in silicon truncated nanocones. The Kerker effect became more commonly used recently due to the explosive growth of dielectric nanophotonics. In this regard, it becomes necessary to study this type of effects considering different scatterer geometries.

The Kerker effect is a unidirectional forward or backward scattering (Conventional and Generalized Kerker effects) or enhanced side scattering (Transverse Kerker effect). The multipole decomposition can serve as an excellent tool to study the Kerker effect. It becomes possible to show the electric field as the sum of multipole contributions, where each term is responsible for the field of a particular multipole[18].

### 2.1. Generalized Kerker conditions

For the first time, the Kerker effect was discovered for spherical particles with the dielectric constant equal to the magnetic permeability. In such particles, when the electric field amplitudes of the electric and magnetic dipoles are equal, as well as at a certain phase difference of the dipoles, one can observe only forward or backward scattering[19]. There are many works that show the experimental realization of such effect in dielectric nanoparticles[52–54]. Over time, other meaningful combinations of multipoles and the phase difference between them were discovered, so the Kerker effect had to be "generalized"[23,55]. Nowadays, the term "generalized Kerker effect" is used, when it comes to pronounced forward or backward scattering.

#### 2.1.1 Amplitude ratio for Generalized Kerker conditions

To describe forward or backward scattering we can write the scattered field of an arbitrary shaped particle under x-polarized light[29], inserting n = (0, 0, $n_z$ ) in (5), we obtain (9):

$$E_x^{sc} = \frac{k^2}{4\pi\varepsilon_0}\frac{e^{ikr}}{r}\left\{ p_x n_z^2 + \frac{1}{c} m_y n_z - \frac{ik}{6} Q_{xz}^e n_z^3 - \frac{ik}{6c} Q_{yz}^m n_z^2 + \cdots \right\} \quad (9)$$

The direction of forward or backward scattering can be linked to the unit vector **n**. Let us set **n** = (0, 0, 1) for the forward scattering, and **n** = (0, 0, −1) for the backward. We consider the case of a plane wave illumination with x-polarization.

Let us assume that we do not have backscattering, and all the multipole contributions that do not participate in the Kerker effect tend to zero. In such cases, the following expressions are valid:

ED + MD: $$E_x^{bwsc} = \frac{k^2}{4\pi\varepsilon_0}\frac{e^{ikr}}{r}\left\{ p_x - \frac{1}{c} m_y \right\} = 0 \quad (10)$$

ED + EQ: $$E_x^{bwsc} = \frac{k^2}{4\pi\varepsilon_0}\frac{e^{ikr}}{r}\left\{p_x + \frac{ik}{6}Q_{xz}^e\right\} = 0 \tag{11}$$

MD + MQ: $$E_x^{bwsc} = \frac{k^2}{4\pi\varepsilon_0}\frac{e^{ikr}}{r}\left\{-\frac{1}{c}m_y - \frac{ik}{6c}Q_{yz}^m\right\} = 0 \tag{12}$$

EQ + MQ: $$E_x^{bwsc} = \frac{k^2}{4\pi\varepsilon_0}\frac{e^{ikr}}{r}\left\{\frac{ik}{6}Q_{xz}^e - \frac{ik}{6c}Q_{yz}^m\right\} = 0 \tag{13}$$

ED + MD + EQ + MQ: $$E_x^{bwsc} = \frac{k^2}{4\pi\varepsilon_0}\frac{e^{ikr}}{r}\left\{p_x - \frac{1}{c}m_y + \frac{ik}{6}Q_{xz}^e - \frac{ik}{6c}Q_{yz}^m\right\} = 0 \tag{14}$$

It should be emphasized that the last combination (14) includes all multipoles up to quadrupoles, and therefore is a superposition of the previous cases. Also, it is worth noting that combination of four multipoles can lead not only to the generalized Kerker effect, but also to the Transverse Kerker effect. Such multipole combinations will be discussed in section 2.3.

Knowing the multipoles values for 'no backscattering' cases, we can express the forward scattering amplitudes:

ED + MD: $$E_x^{fwsc} = \frac{k^2}{4\pi\varepsilon_0}\frac{e^{ikr}}{r}\left\{p_x + \frac{1}{c}m_y\right\} = \frac{2}{c}m_y \tag{15}$$

ED + EQ: $$E_x^{fwsc} = \frac{k^2}{4\pi\varepsilon_0}\frac{e^{ikr}}{r}\left\{p_x - \frac{ik}{6}Q_{xz}^e\right\} = -\frac{ik}{3}Q_{xz}^e \tag{16}$$

MD + MQ: $$E_x^{fwsc} = \frac{k^2}{4\pi\varepsilon_0}\frac{e^{ikr}}{r}\left\{\frac{1}{c}m_y - \frac{ik}{6c}Q_{yz}^m\right\} = -\frac{ik}{3c}Q_{yz}^m \tag{17}$$

EQ + MQ: $$E_x^{fwsc} = \frac{k^2}{4\pi\varepsilon_0}\frac{e^{ikr}}{r}\left\{-\frac{ik}{6}Q_{xz}^e - \frac{ik}{6c}Q_{yz}^m\right\} = -\frac{ik}{3c}Q_{yz}^m \tag{18}$$

Further, considering the equations (12, 4-11), we obtain the ratios between the multipole contributions for each combination of moments leading to the Kerker effect (see Supplementary 2):

ED + MD: $$\frac{C_{sca}^{ED}}{C_{sca}^{MD}} = 1 \tag{19}$$

ED + EQ: $$\frac{C_{sca}^{ED}}{C_{sca}^{EQ}} = 1.67 \tag{20}$$

MD + MQ: $$\frac{C_{sca}^{MD}}{C_{sca}^{MQ}} = 1.67 \tag{21}$$

EQ + MQ:
$$\frac{C_{sca}^{EQ}}{C_{sca}^{MQ}} = 1 \quad (22)$$

### 2.1.2 Phase difference for Generalized Kerker conditions

Another important Kerker effect condition is the phase difference between the interacting multipoles.

For the original problem of Mie scattering on spherical particles, it was shown that electric and magnetic multipoles of the same order have opposite parity with respect to cos θ (θ is the scattering angle with respect to the forward direction)[56]. This means that simultaneous forward scattering amplification and backscattering suppression can be achieved not only for ED and MD interference, but also for any higher-order multipoles, and scattering can be further enhanced or suppressed[57].

The fact is that the phase will directly affect the shape of the scattering diagram we obtain. For all the above combinations of multipoles, specified phase differences can be found to ensure forward scattering.

ED + MD:
$$\Delta\varphi(p,m) = \arg(p) - \arg(m) = \pi \pm 2\pi n, n \in Z \quad (23)$$

ED + EQ:
$$\Delta\varphi(p,Q^e) = \arg(p) - \arg(Q^e) = \pi \pm 2\pi n, n \in Z \quad (24)$$

MD + MQ:
$$\Delta\varphi(m,Q^m) = \arg(m) - \arg(Q^m) = \pi \pm 2\pi n, n \in Z \quad (25)$$

EQ + MQ:
$$\Delta\varphi(Q^e,Q^m) = \arg(Q^e) - \arg(Q^m) = \pi \pm 2\pi n, n \in Z \quad (26)$$

In our work, all the Generalized Kerker conditions with forward direction were obtained for truncated conical nanoscatterers (Figure 3).

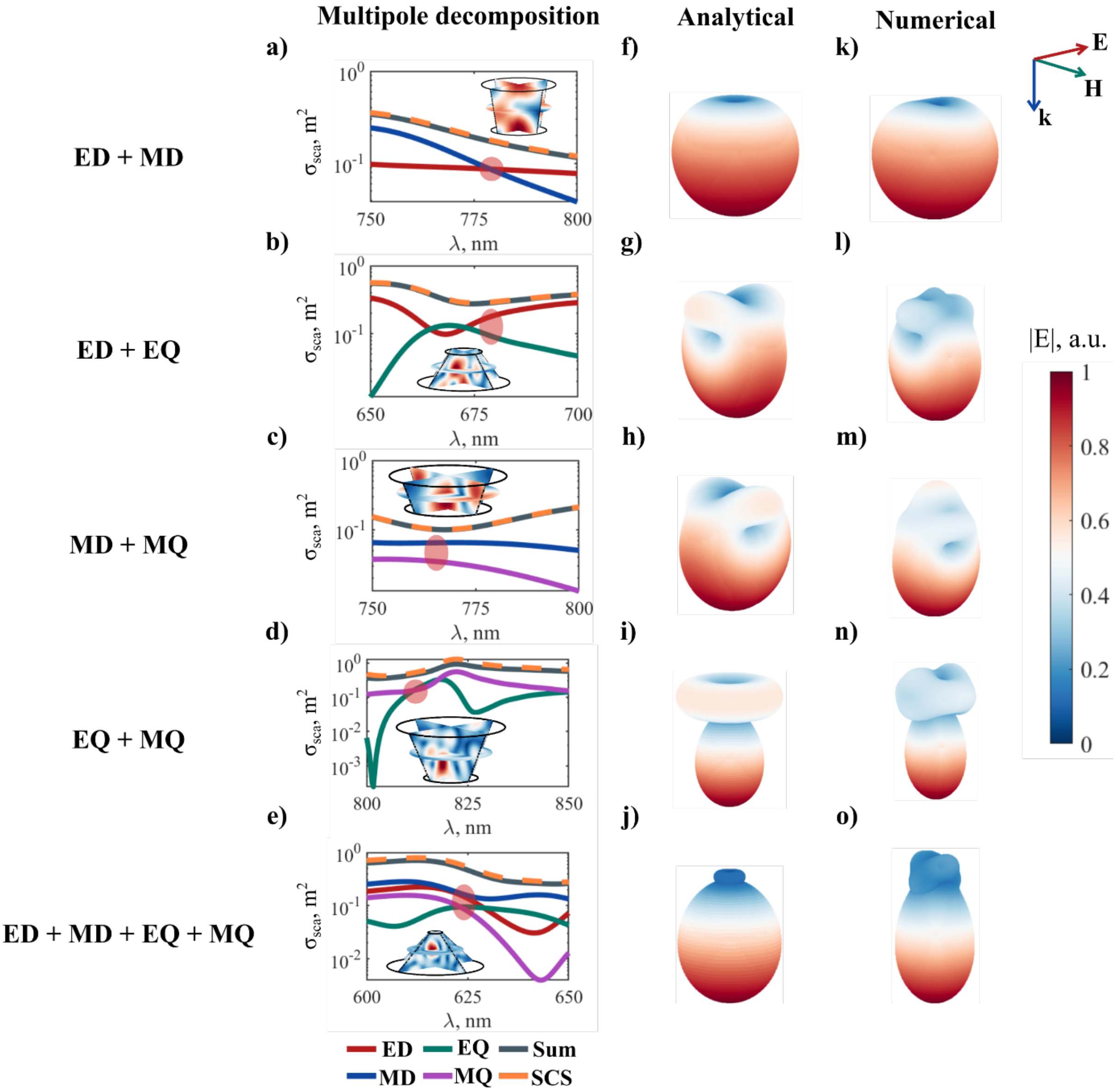

**Figure 3.** Generalized Kerker effects for truncated conical nanoparticles with different shapes. (a-e) – multipolar decomposition, (f-j) - far-field sum of multipoles for a point particle, calculated analytically, (k-o) - far-field distribution for conical nanoparticles with different geometries calculated numerically for real particles at the spectral points shown in (a-e). Conditions of the illumination – incident plane wave from the top of the page. The geometrical parameters for every case can be found in section S3 of the Supplementary Information.

Figure 3 shows the Kerker effects calculated numerically and analytically. The dependence of the multipole contributions to the scattering cross-section on the radiation wavelength is shown in Fig. 3 (a-e). The points of the Kerker effect are marked in red, following the equations (19) - (22). To confirm this, the analytical (f-j) and numerical (k-o) far-field patterns are shown for each case. The scattering shape is similar but not completely the same due to the minor contributions of other multipoles in the numerical calculations.

Thus, various kinds of Kerker effects, and necessary far field and scattering cross-section conditions are shown for real silicon truncated cones. The resulting cases are in a good agreement with the expected

Kerker-type effects for an ideal point calculations. The absence of the need to use higher-order multipoles is also evidenced by good coincidence of the sum of multipoles with numerical calculation.

## 2.2. Transverse Kerker conditions.

Another equally important scattering feature is the transverse Kerker effect, which was first described in[27,58]. The main cause of the transverse Kerker effect is the combination of multipoles featuring scattering to the sides only leaving the small portion of forward scattering according to optical theorem[59]. Such effect can be obtained both through simple combinations of two multipoles[60] and through more complex configurations[28].

Let us assume that both backward and forward scattering are suppressed, and all the multipole contributions not participating in the Transverse Kerker effect tend to zero. Then, the following system of equations is valid for possible ways to obtain Transverse Kerker effect (4 multipoles and 2 multipoles):

ED + MQ:
$$E_x^{bwsc} = E_x^{fwsc} = \frac{k^2}{4\pi\varepsilon_0}\frac{e^{ikr}}{r}\left\{p_x - \frac{ik}{6c}Q_{yz}^m\right\} = 0 \tag{27}$$

MD + EQ:
$$\begin{cases} E_x^{bwsc} = \dfrac{k^2}{4\pi\varepsilon_0}\dfrac{e^{ikr}}{r}\left\{-\dfrac{1}{c}m_y + \dfrac{ik}{6}Q_{xz}^e\right\} = 0 \\ E_x^{fwsc} = \dfrac{k^2}{4\pi\varepsilon_0}\dfrac{e^{ikr}}{r}\left\{\dfrac{1}{c}m_y - \dfrac{ik}{6}Q_{xz}^e\right\} = 0 \end{cases} \tag{28}$$

ED + MD + EQ + MQ:
$$\begin{cases} E_x^{bwsc} = \dfrac{k^2}{4\pi\varepsilon_0}\dfrac{e^{ikr}}{r}\left\{p_x - \dfrac{1}{c}m_y + \dfrac{ik}{6}Q_{xz}^e - \dfrac{ik}{6c}Q_{yz}^m\right\} = 0 \\ E_x^{fwsc} = \dfrac{k^2}{4\pi\varepsilon_0}\dfrac{e^{ikr}}{r}\left\{p_x + \dfrac{1}{c}m_y - \dfrac{ik}{6}Q_{xz}^e - \dfrac{ik}{6c}Q_{yz}^m\right\} = 0 \end{cases} \tag{29}$$

After some simple algebra, one can obtain the following conditions for the multipole amplitudes:

ED + MQ:
$$p_x = \frac{ik}{6c}Q_{yz}^m \tag{30}$$

MD + EQ:
$$\frac{1}{c}m_y = \frac{ik}{6}Q_{xz}^e \tag{31}$$

ED + MD + EQ + MQ:
$$\begin{cases} p_x = \dfrac{ik}{6c}Q_{yz}^m \\ \dfrac{1}{c}m_y = \dfrac{ik}{6}Q_{xz}^e \end{cases} \tag{32}$$

Further, taking into account the equations (27) - (29), we obtain the ratios between the multipole contributions for each of their combinations (see Supplementary S2):

ED + MQ: $$\frac{C_{sca}^{ED}}{C_{sca}^{MQ}} = 1.67 \quad (33)$$

MD + EQ: $$\frac{C_{sca}^{MD}}{C_{sca}^{EQ}} = 1.67 \quad (34)$$

ED + MD + EQ + MQ: $$\begin{cases} \frac{C_{sca}^{ED}}{C_{sca}^{MQ}} = 1.67 \\ \frac{C_{sca}^{MD}}{C_{sca}^{EQ}} = 1.67 \end{cases} \quad (35)$$

For all the above combinations, phase differences between multipoles can be obtained.

Transverse phase conditions are:

ED + MQ: $$\Delta\varphi(p,Q^m) = \arg(p) - \arg(Q^m) = \pi \pm 2\pi n, n \in Z \quad (36)$$

MD + EQ: $$\Delta\varphi(m,Q^e) = \arg(m) - \arg(Q^e) = \pi \pm 2\pi n, n \in Z \quad (37)$$

ED + MD + EQ + MQ: $$\begin{cases} \Delta\varphi(p,Q^m) = \arg(p) - \arg(Q^m) = \pi \pm 2\pi n, n \in Z \\ \Delta\varphi(m,Q^e) = \arg(m) - \arg(Q^e) = \pi \pm 2\pi n, n \in Z \end{cases} \quad (38)$$

In our work, all the Transverse Kerker conditions are shown for truncated conical nano-scatterers (Figure 4).

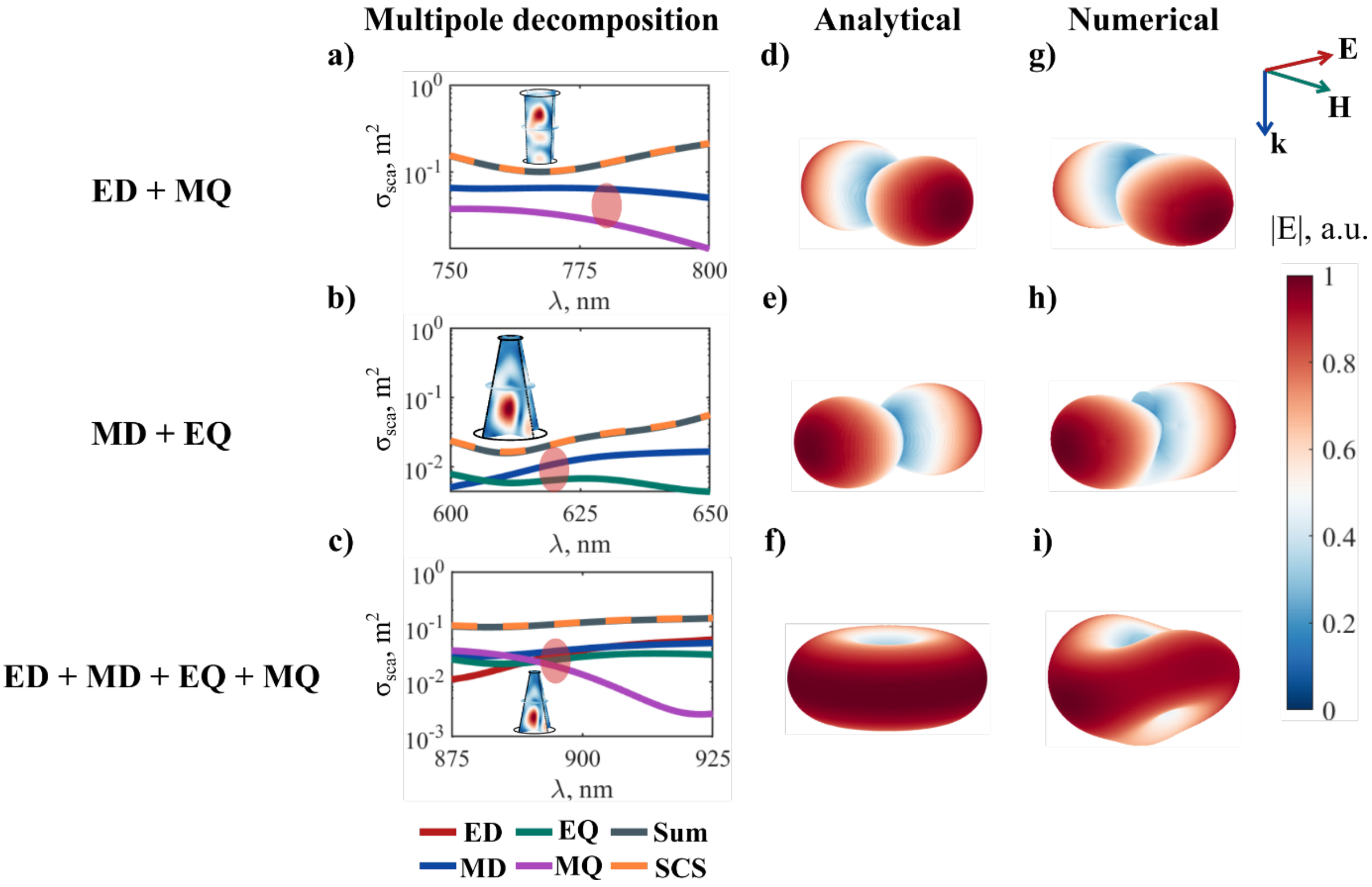

**Figure 4.** Transverse Kerker effects for truncated conical nanoparticles with different shapes. (a-c) – multipolar decomposition, (d-f) - far-field sum of multipoles for a point particle, calculated analytically, (j-g) - far-field distribution for conical nanoparticles with different geometries calculated numerically for real particles at the spectral points shown in (a-e). Conditions of the illumination – incident plane wave from the top of the page. The geometrical parameters for every case can be found in section S3 of the Supplementary Information.

Figure 4 shows the different transverse Kerker effects obtained both numerically and analytically. Scattering patterns shape is an obvious reason why this effect is called "transverse". By the proper choice of the multipole phases and amplitudes, it is possible to achieve side scattering together with forward and backward scattering suppression.

Kerker effects are indeed unique phenomena, which are key to a range of novel photonic devices. Prior to our work, each one of these had previously been only shown for completely different shapes and materials of nano-scatterers. The following are some examples demonstrating various Kerker-type effects.

| | **Generalized** | **Transverse** |
|---|---|---|
| **ED + MD** | Silicon spherical nanoparticles[53,61]. | - |
| **ED + EQ** | Plasmonic gold nanoring nanoantenna[23]. | - |
| **ED + MQ** | - | Ceramic spheroidal particle[62]. |
| | | Silicon square nanoplate[60]. |
| **MD + EQ** | - | Core-shell SiO2@InSb and Si@InSb in a one-dimensional (1D) metalattice geometry in the Terahertz range[63]. |
| **MD + MQ** | Ceramic core-shell[64]. | - |
| **EQ + MQ** | Both isolated and periodically arranged homogeneous cross dielectric structure[65]. | - |
| **ED + MD + EQ + MQ** | Individual core-shell nanoparticles[66]. | Silicon cube[27].<br>Ceramic cube and cylinder[28]. |

**Table 1.** Examples demonstrating various Kerker-type effects for different shapes and material of scatterers.

In this section, we have demonstrated for the first time all known Kerker effects for single nanoscatterers within the same geometry. Thus, nanocones represent a versatile, fabrication-friendly platform for the implementation of new photonic devices benefitting from a comprehensive toolbox of multipolar interference effects including flexible tailoring of scattering patterns.

## 3. Non-scattering regimes: anapole and hybrid anapole

In the past few years, the emergent field of 'anapole electrodynamics' is experiencing exponential growth[67]. Anapoles are semi-nonradiating sources that arise due to the destructive interference of the quasistatic electric dipole moment and the toroidal dipole in the far field. Alternatively, in a more general picture, they can be understood as being originated by the destructive interference of symmetry-compatible quasinormal modes[31]. However, the energy stored by the quasinormal modes within the nanoparticle is nonzero, leading to counterintuitive light-matter interaction processes in the absence of elastic scattering. Until very recently, the suppression of scattering in these states was limited to the electric dipole contribution to radiation. The experimental demonstration of *hybrid anapoles* (HA) following the pioneering theoretical proposal[68], has evidenced the possibility to simultaneously overlap the zeros of all the dominant multipolar channels through a careful design of the nanoparticle geometry. These novel states are much more promising than their dipolar counterparts for a number of reasons[31,32]; despite the larger volume of the nanoresonator required to obtain them, the scattering suppression is improved by more than 20 times, while the excited quasinormal modes store approximately 10 times more energy. Such values exceed by far the performance of anapoles and 2$^{nd}$ order anapoles[69] arising in homogeneous disk nanoresonators.

In nanocylinders pertaining to the symmetry group $D_{\infty h}$, it was shown in[31] how the anapoles from different multipolar orders but equal parity were connected. Counterintuitively, such connection allows to simultaneously overlap four anapoles with just the two geometrical degrees of freedom of the nanoparticle. As mentioned earlier, however, the even and odd multipoles are no longer necessarily coupled once the reflection symmetry in z is broken. In this section, we aim at investigating the effect of z-symmetry breaking on a HA nanocylinder, by introducing a small geometrical perturbation on the upper radius. Such a situation occurs quite often, (unintentionally) during a sample fabrication process.

It is rarely possible to obtain particles of an ideal shape, and samples with small defects can often be found. In practice, the manufactured nanocylinders are most likely truncated cones[70,71].

Figure 5a displays the exact multipole decomposition of the HA regimes of the silicon cylinder studied in[31,32]. The simultaneous suppression of the ED, MD, EQ, and MQ channels at the same spectral position can be appreciated. The inset depicts the norm of the electric field at the spectral point with the lowest scattering, hereafter referred to as the 'HA wavelength ($\lambda_{HA}$)'. The field can be seen to be strongly concentrated within the nanoparticle, (where two hotspots appear), with the exception of a few hotspots at the surface with a size of the order of $\lambda_{HA}/40$. Thus, even in the near field outside the particle, the incident wave is barely distorted by scattering. We now keep the height of the resonator constant and progressively decrease $R_{top} / R_{bottom}$. The multipole decompositions for three selected conicities are shown in Figure 5a-c . In all cases, illumination from the top is considered. A log scale is used to enhance the contrast of the zeros in the spectrum.

Firstly, we observe a spectral blue-shift of all resonant features. This is expected, due to the overall size reduction of the lateral dimensions of the resonator. Secondly, the original MQ and ED anapoles blue-shift faster than the EQ and MD. As a result, the four anapoles are no longer superposed in the spectrum, and the scatterer becomes less 'transparent' to the incident illumination. This is demonstrated by calculations of the total SCS as a function of conicity (Figure 5d).

However, we also observe a progressive emergence of *new anapoles* in the ED and the MQ SCSs (red and purple solid lines, respectively), coinciding with the maxima in the EQ and MD SCSs near the HA. These new features become more pronounced with decreasing $R_{top} / R_{bottom}$. Interestingly, their formation is acompanied by a 'split' (two separated anapoles close to each other, see Figure 5c) in the unperturbed ED and MQ resonances.

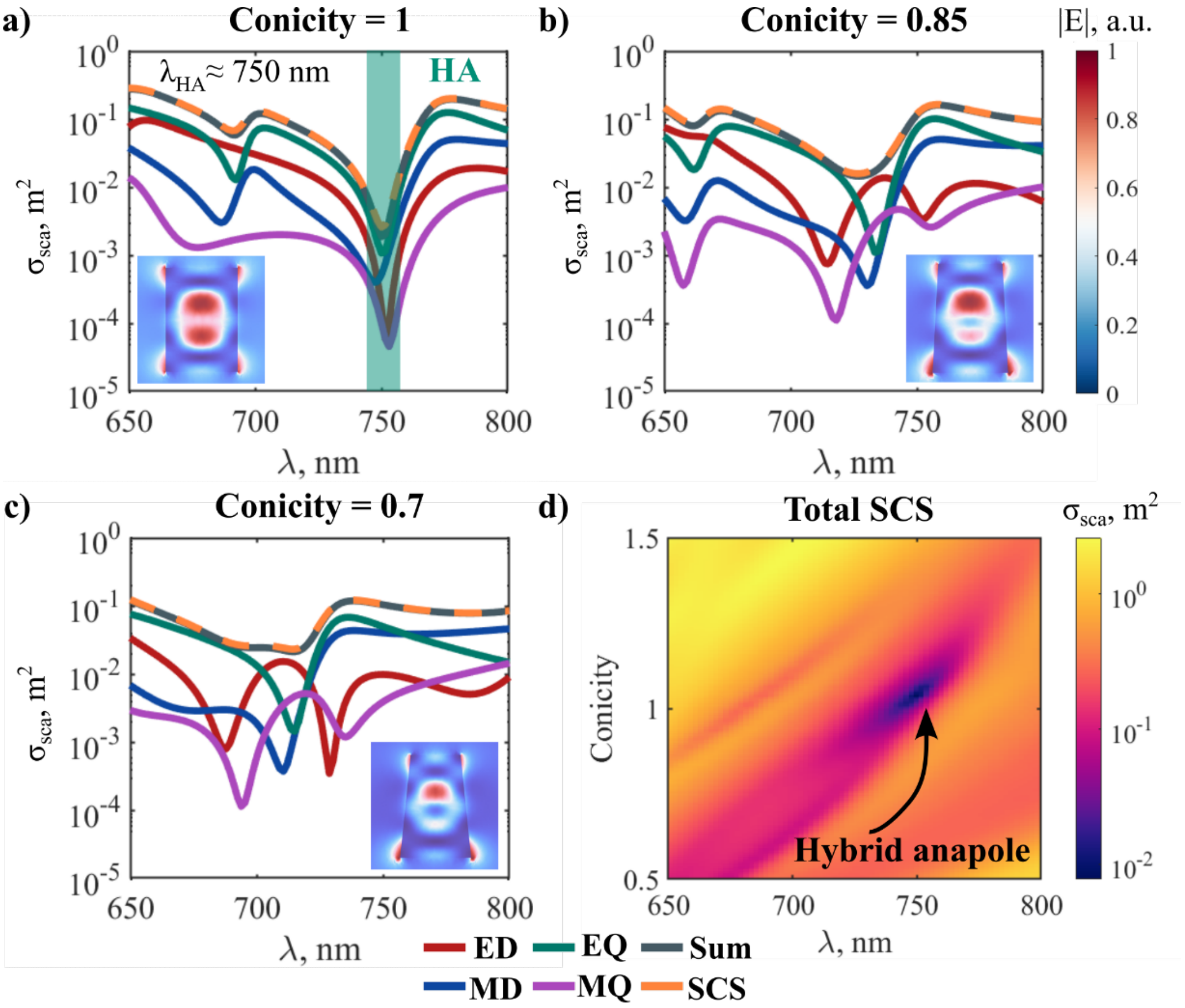

**Figure 5.** Evolution of the HA as a function of conicity and wavelength. (a-c): Multipole decompositions of the SCS (semilogarithmic scale) for selected conicities (see geometric parameters of HA in section S3 of the Supplementary Information). In all calculations, the height of the nanoparticle was kept constant. Insets: distribution of the electric field norm at $\lambda_{HA}$ (d): Total SCS as a function of conicity and wavelength.

We evaluate quantitatively this effect by calculating the wavelength shifts of the multipolar anapoles as a function of conicity (Figure 6). The white-dashed lines show the paths followed in parameter space (only the anapoles involved in the vicinity of the HA regime are investigated). The unperturbed anapoles in the ED and MQ channels are labeled as $A_1$ and $B_1$, respectively (Figure 6a, d). As could be anticipated from the results in Ref.[31], the latter follow the same path in parameter space. This is because they are both associated to the same resonant QNM. The same occurs with the MD and EQ anapoles (Figure 6b, c).

For values of $R_{top} / R_{bottom} < 1$, we confirm the appearance of new anapoles in both the ED and MQ channels, manifesting as pronounced dips in their contributions to the SCS (Figure 6a, d). For convenience, we denote them as $A_2$ and $B_2$. Importantly, we notice that they both follow a similar path in parameter space. What's more, there seems to be a connection between the paths of $A_2$ and $B_2$ and those followed by the original EQ and MD anapoles; after their appearance, the former spectrally overlap with the latter. In contrast, no appreciable new features can be seen when $R_{top} / R_{bottom} > 1$. In a first regard, this observation seems contradictory, since symmetry is analogically broken. However, later the physical reason for it will become clear. It is necessary to mention that in the calculations of the conicity we keep $R_{bottom}$ constant, then the volume of the $R_{top} / R_{bottom} > 1$ is larger than $R_{top} / R_{bottom} < 1$.

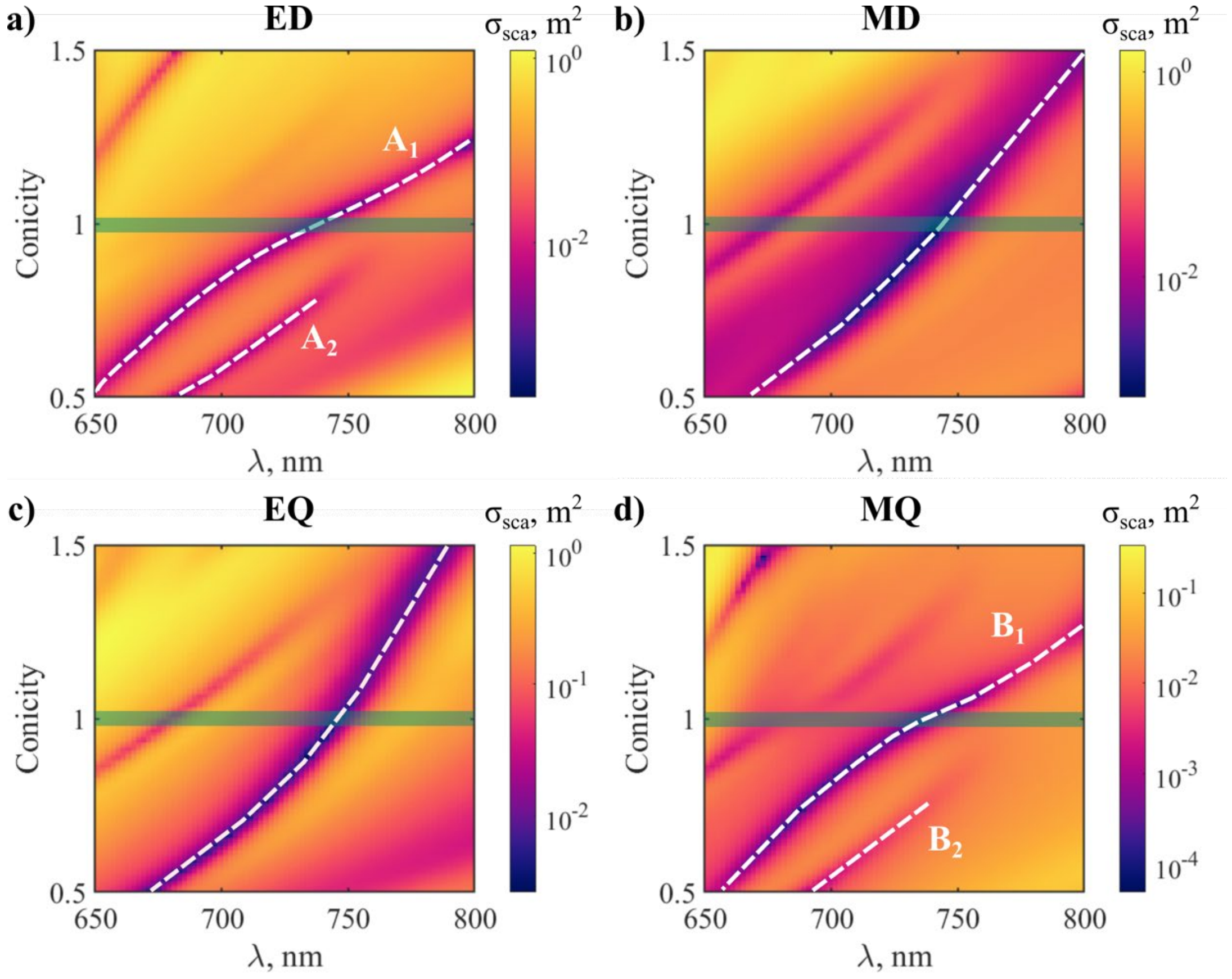

**Figure 6.** Evolution of the SCS of each multipole (in log scale), as a function of wavelength and conicity. White dashed lines are a guide to the eye, and indicate the paths traced by anapoles in parameter space. The shaded green lines highlight the unperturbed cylinder (with conicity = 1). (a) ED. $A_1$ labels the original ED anapole in the unperturbed cylinder, and $A_2$ is the new ED anapole induced by symmetry breaking. (b) MD, (c) EQ, (d) MQ. $B_1$ labels the original MQ anapole in the unperturbed cylinder, and $B_2$ is the new MQ anapole induced by symmetry breaking.

To gain more insight, we briefly revisit the origin of anapoles, from the QNM perspective[72,73]. As an example, consider a scatterer supporting one resonant QNM radiating to the ED channel. The SCS would then be well approximated by the first term in equation (6), i.e.:

$$\sigma_{sca} \propto |\mathbf{p}|^2 = |\mathbf{p}_{bg} + \mathbf{p}_1|^2 . \tag{39}$$

Here, $\mathbf{p}_1$ is the induced ED moment by the resonant QNM, and $\mathbf{p}_{bg}$ corresponds to a non-resonant contribution stemming from QNMs outside the spectral range of interest, as well as direct scattering from the object's shape[74]. A zero (anapole), takes place when $|\mathbf{p}|^2 = 0$, so that the resonant contribution cancels out with the background:

$$\mathbf{p}_1 = -\mathbf{p}_{bg} . \tag{40}$$

We note that, in general, $\mathbf{p}_{bg}$ is a smooth function of frequency that cannot, a priori, be controlled by design, while $\mathbf{p}_1$ becomes non-negligible only near the resonance frequency. In particular, one can model $\mathbf{p}_1$ as a Lorentzian function centered at the resonance frequency. In the presence of a second resonant QNM, equation (40) can be expressed as:

$$\mathbf{p}_1 + \mathbf{p}_2 = -\mathbf{p}_{bg} . \quad (41)$$

Equation (41) provides one more degree of freedom to reach the anapole condition. In principle, it is possible to tune $\mathbf{p}_1$ so that $\mathbf{p}_1 = -\mathbf{p}_{bg} - \mathbf{p}_2$, or change $\mathbf{p}_2$ so that $\mathbf{p}_2 = -\mathbf{p}_{bg} - \mathbf{p}_1$. Since $\mathbf{p}_{1,2}$ vary naturally in amplitude and phase near the resonance frequencies of the associated QNMs, a new anapole should emerge for every new QNM contributing to the ED cross-section[75]. This conclusion can be readily generalized to any multipole.

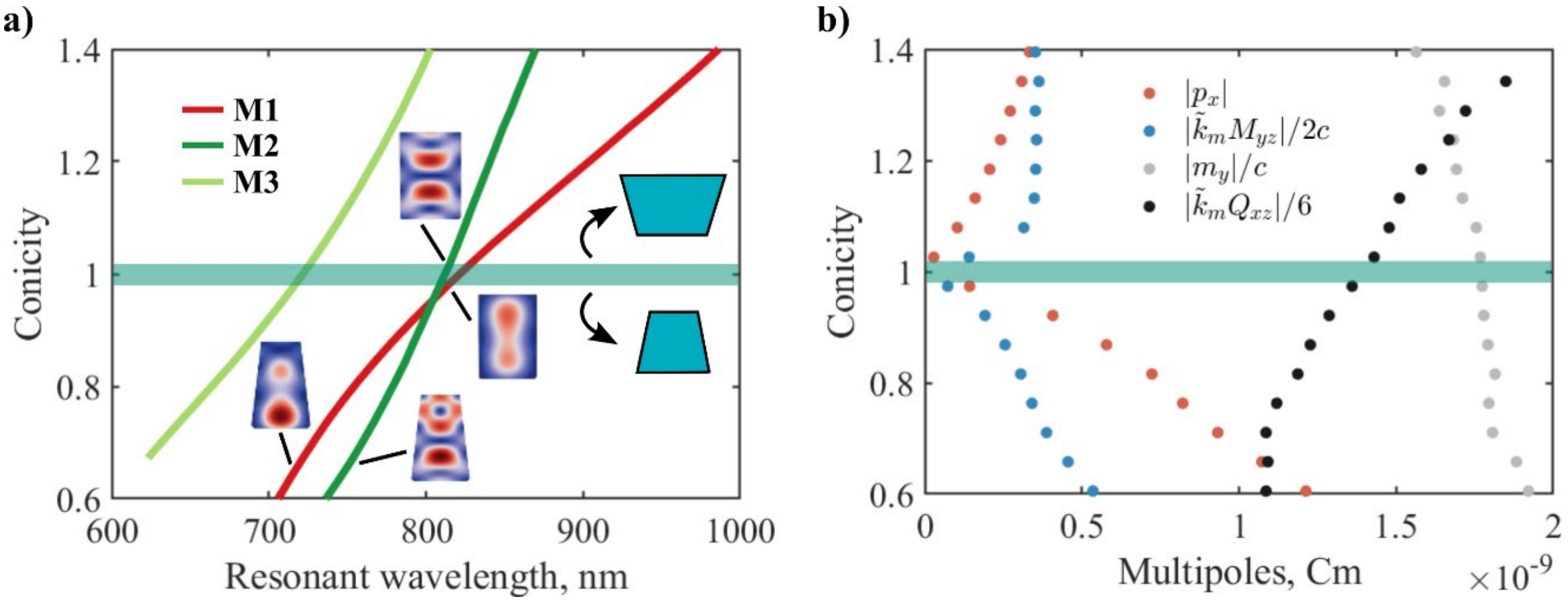

**Figure 7.** (a) Resonant wavelength shifts of the QNMs in the vicinity of the HA, as a function of conicity. In an unperturbed cylinder, QNMs M1 and M2 correspond to $\mathrm{TE}_{120}$ and $\mathrm{TM}_{113}$ discussed in Ref.[31]. M3 is associated with $TM_{111}$. Insets depict the internal fields of the QNMs for selected values of conicity (red – more intense field, blue – less intense field). In all the calculations we neglected the dispersion of Si, and fixed its value to an average in the spectral range under consideration, $n \approx 3.45$. (b) Intrinsic multipole moments of M2 as a function of conicity. To allow a proper visualization, all moments are normalized in ED units, i.e. C.m. $\tilde{k}_m = \tilde{\omega}_m / c$ is the (complex) wavevector associated with the m-th QNM.

From the above discussion, it follows that anapole formation is intrinsically related to modal evolution. In particular, the emergence of a new anapole in the ED and the MQ SCSs must be directly connected to an additional QNM contribution to these multipoles. With this idea in mind, we calculate the resonant QNMs in the vicinity of the HA as a function of the conicity (Figure 7a). Three QNMs were considered: the two QNMs primarily responsible for the HA effect (M1 and M2 in Figure 7a) and, for completeness, a third QNM lying within the visible range (M3 in Figure 7a).

In Ref.[31], it was shown that M1 was associated with the formation of both the ED and MQ anapoles ($A_1$ and $B_1$), while M2 was responsible for the MD and EQ ones. This is further confirmed here by comparing the paths of the anapoles in Figure 6 with the evolution of the resonant wavelengths in Figure 7a. The role of M3 can be safely neglected in the discussion (or considered part of the background), since it is spectrally isolated from M1 and M2 in the range of parameters considered.

We now focus our attention on the multipolar character of QNM M2 (Figure 7b). We calculate the intrinsic multipole moments as explained in section 1.2. This indicates us whether a QNM will contribute or not to the SCS of a given multipole. For our purposes, if the contribution of a specific QNM is zero, it cannot play a role in the formation of an anapole in that multipole SCS.

The results reveal a surprising fact: for an unperturbed cylinder, the contributions of M2 to the ED and MQ are zero, while the MD and EQ are not, as expected since it interferes in those two channels to form anapoles. However, when $R_{top} / R_{bottom} \neq 1$, M2 starts contributing to the ED and MQ SCSs. The MQ and ED contents of M2 grow as a function of the perturbation. It immediately follows from the discussion above that a new anapole can emerge in the two multipoles under consideration. Furthermore, the contributions to the MD and EQ SCS also change: for $R_{top} / R_{bottom} > 1$ the QNM is better matched to the EQ. Conversely, for $R_{top} / R_{bottom} < 1$, the MD becomes dominant.

The fact that M2 couples to multipoles with opposite parities when symmetry is broken can be understood from Figure 2 and the discussion in section 1.2. In a truncated cone, the modes are no longer eigenstates of parity, and therefore can scatter light as a mixture of even and odd multipoles.

In brief, the emergence of anapoles $A_2$ and $B_2$ is due to mode M2 being able to contribute to the ED and MQ SCS, once cylindrical symmetry is broken. This also explains the apparent connection between the paths followed by the EQ and MD anapoles (associated with M2) and the new emerging anapoles. The reason why $A_2$ and $B_2$ appear only for $R_{top} / R_{bottom} < 1$ can be elucidated from the behavior of the resonant wavelengths in Figure 7a. It can be seen that M1 and M2 remain spectrally close when decreasing $R_{top} / R_{bottom}$. In contrast, for the range of parameters considered, M1 rapidly redshifts towards the near IR with increasing $R_{top} / R_{bottom}$, while M2 remains in the visible. In consequence, M1 is spectrally isolated from M2, and the latter does not play an important role anymore. Thus, no new anapoles are formed.

Regarding the robustness of the HA regime under small perturbations, extensive numerical tests demonstrate that it is still possible to recover the spectral overlap of the four anapoles up to perturbations in the range of $R_{top} / R_{bottom} = \{0.8, 1.2\}$ by varying the resonator height (not shown). Interestingly, the lower limit appears to coincide with the region where $A_2$ and $B_2$ become more pronounced (Figure 7a,d). This suggests that, once the MQ and ED content of M2 becomes significant, the MQ and ED anapoles can no longer be tuned independently from the EQ and MD ones, which constituted the general strategy to design HA in nanocylinders.

Next, we study the possibility to realize HA in truncated cones with very small $R_{top} / R_{bottom}$ (Figure 8). Keeping the bottom radius and the height constant, we perform calculations of the total SCS for conicities in the range between 0 (perfect cone) and 0.5 (Figure 8a). Our simulations reveal the existence of regions with strongly suppressed SCS (white square in Figure 8a). Figure 8b displays the multipole decomposition for a nanocone with conicity 0.2. A HA can be seen to form due to the overlap between the ED and EQ anapoles, and the spectral proximity of MD and MQ anapoles. For the example at hand, scattering is suppressed by approximately 8 times the average SCS in the visible range. In the wavelength featuring the lowest SCS the scatterer is virtually transparent to the incident plane wave, as demonstrated in the field profile shown in Figure 8c. The incident plane wave is seen to propagate undistorted by the cone. Although significant, the result leaves room for improvement. More intensive multiparameter searches can lead to a closer overlap between the four anapoles, and an even more pronounced scattering suppression.

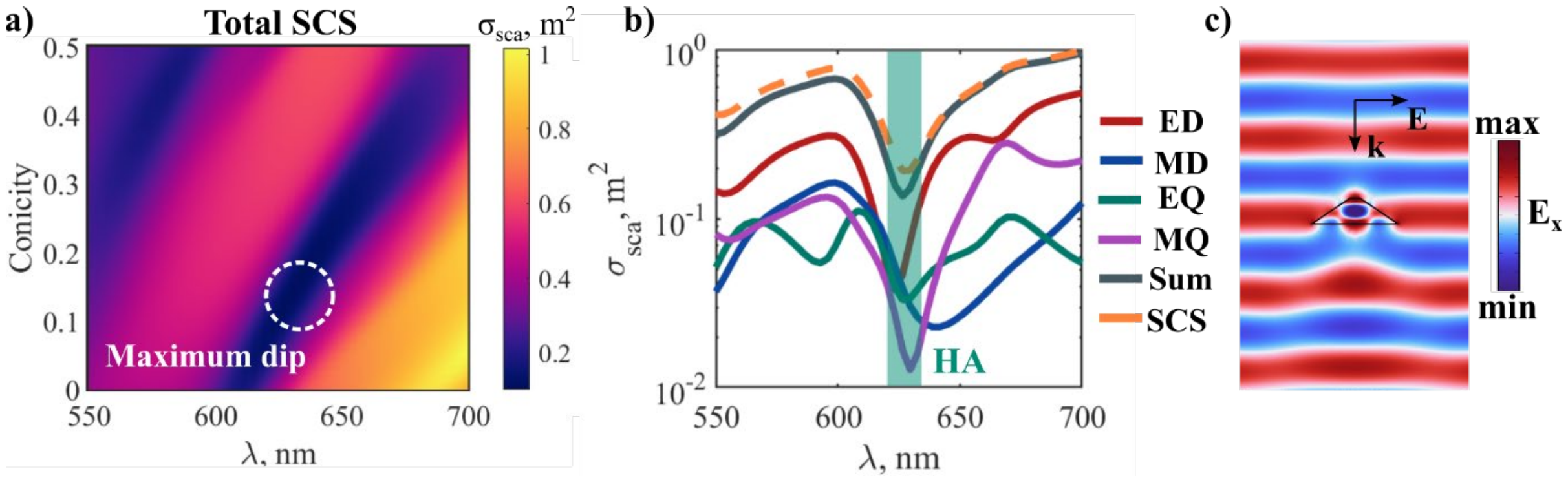

**Figure 8.** (a) Evolution of the total SCS near the HA for large conicities, (nanoparticles with the same height and radius as in Figure 7). The white-dashed circle highlights the region with the strongest scattering suppression in the parameter range considered. (b) Multipole decomposition of the SCS for a truncated nanocone with conicity 0.12, lying within the region of maximum scattering suppression shown in (a). The green-shaded area indicates the spectral range with the maximum scattering suppression (approximately 8 times less than the average SCS in the visible). (c) x-component of the electric field when illuminating the nanocone with an x-polarized plane wave at a wavelength of 625 nm, corresponding to the HA.

The study in this section reveals the strong effect that small perturbations in conicity can have in the resonances responsible for the HA, since they effectively break the underlying symmetry, and render previously closed channels open, which leads to additional interactions between the modes. The knowledge of these can be important for future applications that benefit from the exotic properties of HA, for instance, metasurface engineering, sensing, etc. Furthermore, we have demonstrated, for the first time, the possibility to achieve HA in nanocones. In this way, we have expanded the library of all-dielectric nanostructures that support HA, beyond cylinders[31] and ellipsoids[34]. Once again, cones present themselves as a simple, fabrication-friendly platform for the implementation of multipolar interference effects.

## 4. Superscattering regime

In the last section of this tutorial, we introduce the superscattering effect. In contrast to the hybrid anapole this effect concerns scattering enhancement from a subwavelength particle. This unique regime has already found a plethora of emerging applications, such as, e.g. sensing[76], energy harvesting[77], radar deception[78], etc. To enhance scattering, it is generally believed that one needs to spectrally overlap the resonant frequencies of several QNMs scattering to different multipoles[35]. It follows from a simple geometry tuning where the thicknesses of core-shell structures were optimized to precisely bring modes together. Since the structure retains spherical symmetry, no multipole mixing is allowed, and thus there is no interaction between the different QNMs. Several other works followed on the same foot-steps; the superscattering regime has been investigated in a variety of structures retaining spherical symmetry[37,56,79–81] and was recently confirmed experimentally in the microwave frequency range[82].

The generally accepted definition of superscattering is as follows[35]: the total scattering cross-section of a resonator must exceed (by far) the maximum scattering cross-section of a dipolar particle with spherical symmetry[35]. The maximum contribution of a multipolar channel to the scattering for the specified resonator can be formulated as [35,83],

$$C_{\max}^{l} = \frac{2l+1}{2\pi}\lambda^2, \tag{42}$$

where $l$ is the total angular momentum and $\lambda$ is the wavelength. Thus, for the first order multipoles (electric and magnetic dipoles) the maximum cross-section reduces to

$$C_{\max}^{p} = \frac{3}{2\pi}\lambda^2 , \tag{43}$$

Hereafter we refer to this limitation as the dipolar maximum for the spherically symmetric scatterer or the DM as an abbreviation. Hence, superscattering corresponds to the condition $C_{\max} >> C_{\max}^{p}$.

Recently, it has been noticed that departing from the spherical scatterer to a scatterer lacking a rotation and/or a reflection symmetry, this bound no longer holds[83]. Therefore, the scattering maxima of the multipoles are no longer limited by equation (42). In fact, to our knowledge, in a scatterer of arbitrary size and shape, the contribution of a multipole to the scattering cross-section is not bounded[48]. In the literature, however, the scattering limitations defined for spherical scatterers remain as a benchmark for the superscattering occurrence. Henceforth, we are normalizing the scattering by the dipolar channel maxima defined in equation (43) to establish consistent comparisons for the scattering enhancement in different setups.

In our recent work[83], we have suggested and experimentally demonstrated a new paradigm to achieve superscattering. It benefits from the vastly growing fields of non-Hermitian physics. Indeed, it has been established that subwavelength open-cavity resonators support QNMs that may interact and collectively lead to a plethora of light-matter interactions effects such as enhanced directionality or broadband scattering.

We have shown[83] starting from a spherical resonator supporting two QNMs that radiate independently to two multipolar channels of different order but equal parity, a fine tuning of the vertical and/or the horizontal radius (i.e., deforming the sphere into a spheroid) results in increasing scattering on a certain multipole so that it exceeds the limitation introduced in equation (42). It also proves that breaking the fully symmetric structure, in this case defined by the sphere, into an axisymmetric structure such as the ellipsoid may lead to strong coupling between two QNMs and furthermore allows controlling the multipolar character of the scattered wave (i.e. its far field characteristics). Thus, for the first time the concept of super multipoles was introduced. This newly suggested pathway to superscattering opens uncharted territories for designing resonators that capture incident photons in an unusually large area. Compared to core-shell spherical particles where superscattering can be achieved only through the accidental overlapping of modes, scattering of reduced symmetry resonators can be enhanced both accidentally (spectrally tuning the resonance frequencies of two noninteracting QNMs) or by carefully tailoring the interference of coupled QNMs.

As emphasized in the previous sections, the tuning of the conicity offers more degrees of freedom in controlling the radiation patterns of the QNMs that are unavailable for scatterers with cylindrical symmetry. Lack of the reflection symmetry in conical geometries allows for interference between modes with unequal parities. This additional degree of freedom has an important consequence on realizing super multipoles as we later explain.

Figure 9(a) displays the total scattering cross-section normalized by the DM (equation (43)) as a function of the conicity and the wavelength. To reduce the computational complexity, we again assume dispersionless Si materials similarly to the setup described in the previous section. Cone's geometrical parameters are enlisted in the figure caption, we here selected a subwavelength cone in the visible and near-infrared region as a practical low-loss superscatterer structure. It is apparent that the normalized scattering exceeds by more than double the DM in the specified range. One can also observe that the now presented "superscattering" broadens while tuning conicity with multiple appreciable superscattering peaks. Noteworthily, the normalization factor (the DM) is not a function of the geometry, therefore, one can assume that the value of superscattering can be readily increased by considering factors of geometry along with the operating wavelength range and the material composition of the resonator.

The resonance frequencies of the QNMs in the spectral range of interest are shown in Figure 9(b) as a function of conicity. The QNMs are labeled as $M_1'$ - $M_4'$. It can be noticed that the cylinder (conicity = 1, marked by the shaded green line) supports the conventional magnetic dipole mode M1 at the resonant wavelength 983 nm and a super-ED mode M2 at resonant wavelength 895 nm[83]. The insets in figure 9(b) show the internal field distribution of the two modes, (detailed field profiles inside and in the vicinity of the cylinder are shown in supplementary figure S2). $M_1'$ and $M_2'$ avoid the crossing, and redshift as the cone volume increases (for an increased conicity). We consider these two QNMs as the focus of our investigation to alter and potentially enhance scattering.

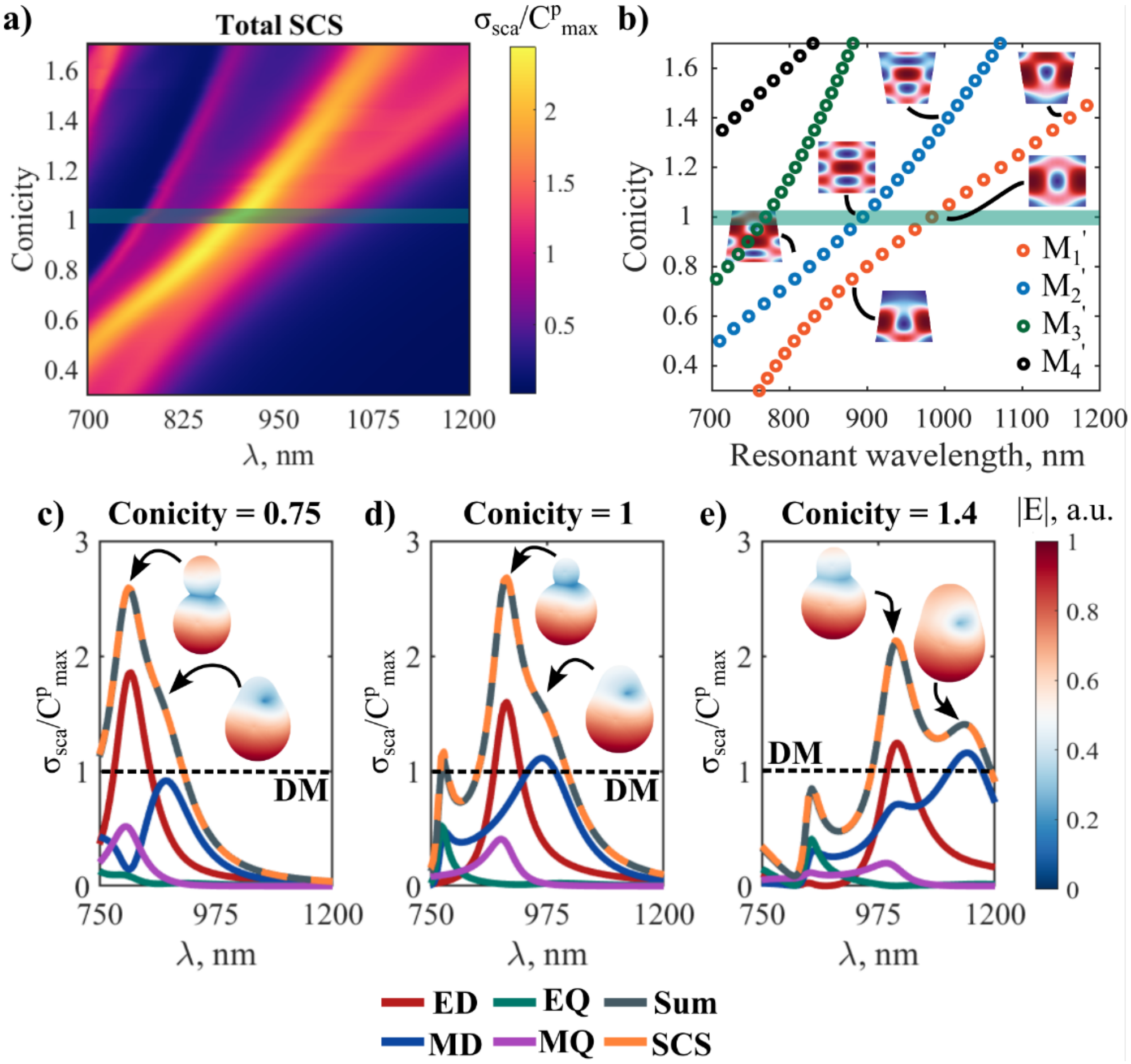

**Figure 9**. Evolution of the superscattering features as a function of conicity and wavelength. (a) Normalized total scattering cross-section of the perturbed cone under normal incidence plane wave excitation. (b) Resonant wavelengths of the QNMs ($M_1'$ - $M_4'$) as a function of the conicity. Insets define modes inner fields at three selected conicities (red - more intense field, blue - less intense field). Cones height and bottom radius considered constant with values (500 nm) and (100 nm), respectively. The cones top radius was modified with constant step size in the range (30 - 170 nm), while the material is considered dispersionless (n = 3.45) for calculations simplicity reasons. (c) Multipole decomposition of the normalized total scattering cross-section for the three selected conicities in (a). Scattering patterns at the peaks of the electric and magnetic dipoles resonances are inserted for each case. The dashed lines

point to the dipolar channel scattering maxima in the case of spherical symmetric resonator (DM). Multipolar channels that contribute more than the DM are superscattering.

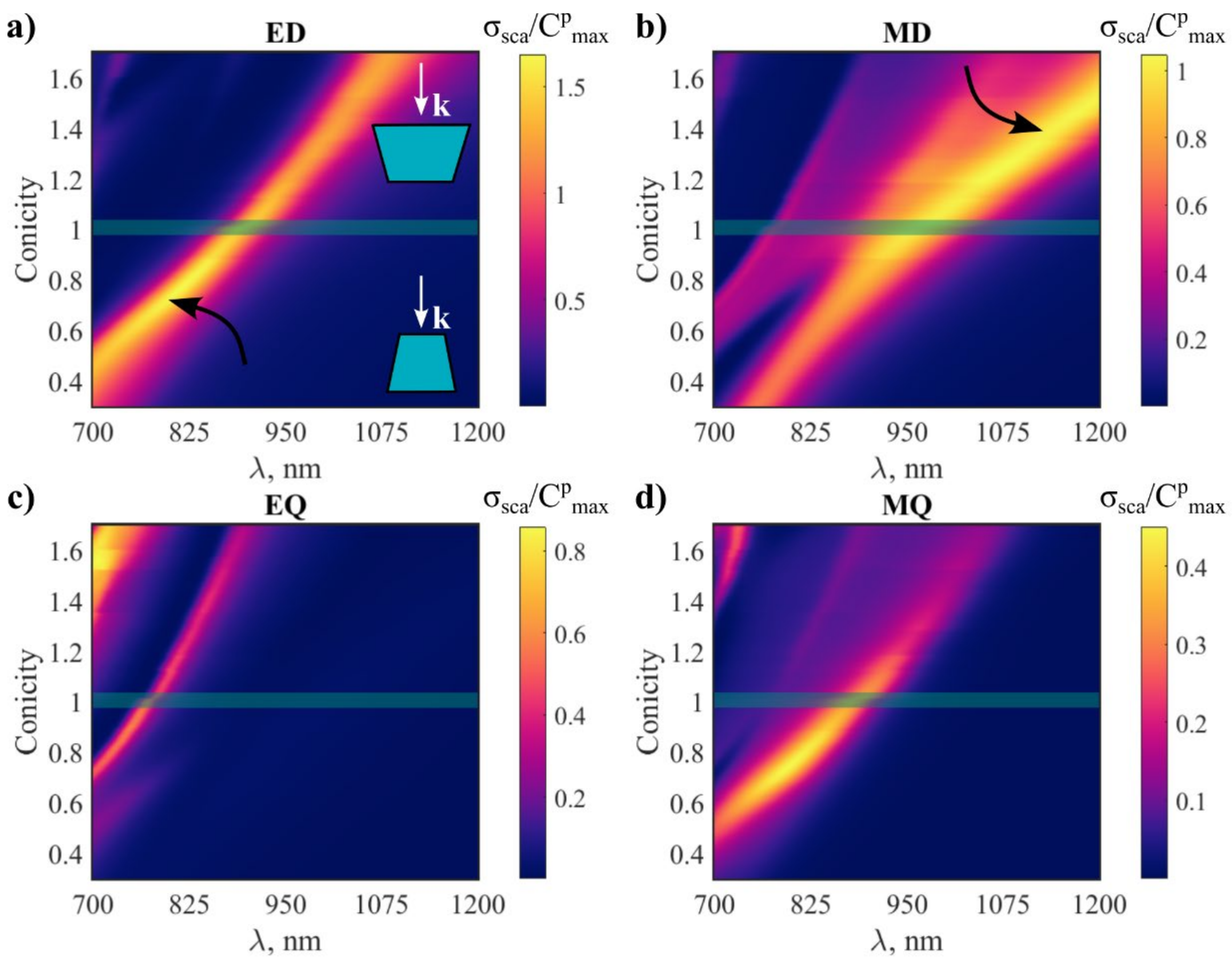

**Figure 10**. The multipole contributions to total scattering of the cone setup presented in Figure 9 with ED (a), MD (b), EQ (c), and MQ(c). The shaded green line corresponds to the unperturbed cylinder (conicity=1). The two black arrows point to the regions of maximum of electric dipole and magnetic dipole contributions in the lower and upper branches of modes M1 and M2 interference.

From figure 9(d), it can be appreciated that the two even multipolar channels, the electric dipole and the magnetic quadrupole, contribute to the scattering at spectral points coinciding with the resonant wavelength of $M_2$'. We notice that the electric dipole is showing superscattering where it exceeds the DM by about 1.5 times. The latter are closely related to the super dipole resonances studied in Ref. [83]. Similarly, $M_1$' and $M_3$' appear to radiate on the odd multipolar channels, the magnetic dipole and the electric quadrupole, respectively. The magnetic dipole displays a scattering cross-section with values slightly above the DM. Summarizing, mode $M_2$' is even and scatters as a combination of electric dipole and magnetic quadrupole moments, while odd QNMs $M_1$' and $M_3$' are odd and scatter as a combination of magnetic dipole and electric quadrupole moments.

We show in figure 9(c) and figure 9(d) cases for conicity 0.75 and 1.4, respectively. In the first, the electric dipole cross-section is now almost double the DM, while the magnetic dipole is reduced below that level. When decreasing the upper radius of the cavity, the resonant wavelengths of QNMs $M_1$' and $M_2$' are blueshifted [Figure 9(b)]. As a result, the multipole resonances are also blueshifted [Figure 9(c)]. The opposite occurs when increasing the upper radius, [figure 9(e)]. However, a second magnetic dipole scattering peak is now observable, overlapping with the electric dipole and magnetic quadrupole peaks. Meanwhile, we show for all the three cases the scattering patterns at the distinguished dipolar scattering peaks. The scattering is more directive when both dipoles are present and comparable in values, and it takes obviously a dipolar scattering signature when either of the dipoles is dominant. To have a clearer picture of the mechanism at play that controls the position and amplitudes of the superscattering multipolar channels, we show in figure 10 the scattering evolution of the first four multipoles as a function of the conicity and the wavelength.

When compared to the QNMs' dispersion in figure 9(b), we notice that the power scattered by the electric dipole is mainly attributed to mode $M_2'$ while it appears that modes $M_1'$ - $M_3'$ are able to scatter through the magnetic dipole channel. For this reason, we see that the magnetic dipole appears with two peaks on the scattering spectrum where it is especially noticeable for higher conicities (conicity > 1.2). Modes $M_3'$ and $M_4'$ across the parametric space are radiating on the electric quadrupole channel. One can notice also that the magnetic quadrupole channel is mostly apparent aligning with the electric dipole peak scattering. Although the quadrupolar channels are not superscattering within themselves in the current design, they have important contributions to the total scattering cross-section.

It can be concluded that modes $M_1'$ and $M_2'$ of the cylinder are allowed to interfere with each other after breaking the vertical mirror symmetry i.e., with conicity $\neq 1$. This implies that both modes now able to radiate on the even and odd parity multipolar channels. One important result of this interference is an increase of the electric dipolar superscattering with almost doubling the DM at small conicities. We can also appreciate that $M_2'$ start to scatter on the magnetic dipolar channel in the bigger conicities region of the parametric figure. Considering this mode is originally even parity in the cylinder, it becomes hybrid when tuning conicity in the sense that it scatters as a mixture of electric and magnetic dipole moments. Therefore, a single QNM in the conical geometry is able to drive opposite parity multipolar channels to superscattering regime. This type of superscattering cannot be obtained in the $D_{\infty h}$ resonators without spectrally overlapping the resonant wavelengths of at least two QNMs. Such hybridization of superscattering channels is particularly important for the superscattering effect since it also decisively impacts the directivity of the overall scattering. Meaning, it can be utilized to enhance forward or the backward scattering upon prerequisites. Another important consequence of $M_1'/M_2'$ interference is the broadening of the superscattering. This can be seen in both the lower and upper regions of the interference.

The above systematic exploration of the superscattering achievable through designing conical geometry has revealed a new approach to obtain desired outcomes in superscattering features. We focused on tuning two modes ($M_1'$ and $M_2'$) of a cylinder that are in spectral proximity, sufficiently confined within the resonator domain, and with an unequal parity. This strategy can be sufficient but is not necessary to achieve superscattering in the geometry under consideration. However, the investigated mechanism highlights important features of superscattering in a cone, namely, (i) the possibility to exceed the DM with the help of the novel supermultipoles and (ii) the mixed character of the involved QNMs, which scatter as a combination of even and odd multipoles, unlike in spheres or cylinders. Alternatively, other strategies remain possible as well, however with optimized conical structure in this work we believe that several important aspects of the superscattering conical geometry have been revealed.

Thus far we have shown a remarkable control over scattering, due to the design flexibility and symmetry rules governing the conical geometry. Tuning a single geometrical parameter (in this case - the conicity) can result in strong coupling between two QNMs of opposite parity, as was previously shown earlier this year[51]. Remarkably, this mechanism can be utilized to enhance scattering and manipulate the radiation pattern at the resonance to render it more directive. It indeed provides a plethora of opportunities to optimize truncated cones for applications in optics.

## Conclusion

In this work, we have proposed the conical geometry as a universal platform for obtaining many important optical effects that are of interest in nanophotonics. For the first time, all known Kerker effects were obtained on one scatterer shape for a real dispersion of the refractive index of silicon. It was found that the recently discovered non-scattering hybrid anapole regime can also be obtained in conical nanoparticles, in addition to elliptical and cylindrical ones. It is shown how, by changing the geometric parameters, it is possible to adjust the anapole regime of different multipoles and obtain a hybrid anapole regime. We also studied the possibility of obtaining superscattering, the effect inverse to the anapole regime, in conical particles, showing the dependence on the geometry of the scatterer. This research takes a step towards nanophotonics of more complex shapes with the ability to fine-tune the effects that can be obtained on a single shape of a nano-scatterer. This research significantly reduces the cost of developing photonic devices and opens up new horizons for the practical application of the next-generation photonics.

## Acknowledgement

The authors gratefully acknowledge the financial support from the Ministry of Science and Higher Education of the Russian Federation (Agreement No. № 075-15-2022-1150). The investigation of the hybrid anapole regimes has been partially supported by the Russian Science Foundation (Grant No. 21-12-00151). V.B. acknowledges the support of the Latvian Council of Science, project: DNSSN, No. lzp-2021/1-0048.

# Supplementary Material for "Special scattering regimes for conical all-dielectric nanoparticles"

Alexey V. Kuznetsov[1,3,*], Adrià Canós Valero[1], Hadi K. Shamkhi[1], Pavel Terekhov[4], Xingjie Ni[4], Vjaceslavs Bobrovs[3], Mikhail Rybin[1], Alexander S. Shalin[2,3*]

[1]ITMO University, Faculty of Physics, St. Petersburg, 197101, Russia

[2]MSU, Faculty of Physics, Moscow, 119991, Russian Federation

[3]Riga Technical University, Institute of Telecommunications, Riga, 1048, Latvia

[4]The Pennsylvania State University, Department of Electrical Engineering, Pennsylvania, 16802, United States

*alexey.kuznetsov98@gmail.com

*alexandesh@gmail.com

## S1. Experimental Dispersion of amorphous Si

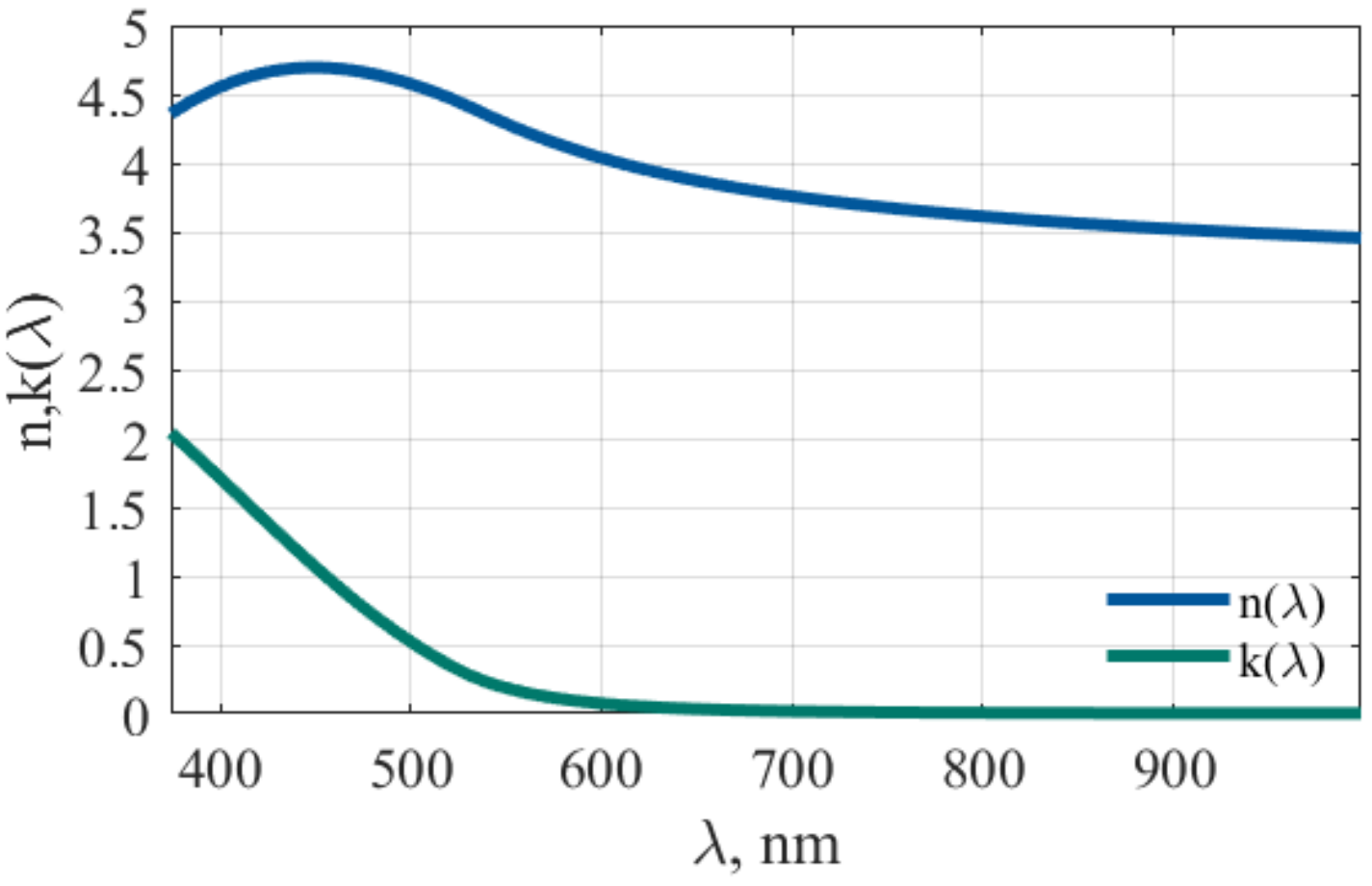

**Supplementary Figure S1.** Real (n) and imaginary (k) parts of the experimentally measured refractive index of amorphous silicon (aSi) employed in our study, as a function of wavelength.

## S2. Conditions for Generalized Kerker effects

In the main part of the article the conditions for the forward Kerker effect were obtained. We used formula (6) in main part of manuscript and vanished the terms that do not participate in the effect.

ED + MD:
$$C_{sca}^{total} = C_{sca}^{p} + C_{sca}^{m} = \frac{k^4}{6\pi\varepsilon_0^2 \left|\mathbf{E}_{inc}\right|^2}\left[\left|p_x\right|^2 + \left|\frac{m_y}{c}\right|^2\right] \quad (1)$$

ED + EQ: $$C_{sca}^{total} = C_{sca}^{p} + C_{sca}^{Q^e} = \frac{k^4}{6\pi\varepsilon_0^2 |\mathbf{E}_{inc}|^2}\left[ |p_x|^2 + \frac{1}{120}\left( |kQ_{xz}^e|^2 + |kQ_{zx}^e|^2 \right) \right] \quad (2)$$

MD + MQ: $$C_{sca}^{total} = C_{sca}^{m} + C_{sca}^{Q^m} = \frac{k^4}{6\pi\varepsilon_0^2 |\mathbf{E}_{inc}|^2}\left[ \left|\frac{m_y}{c}\right|^2 + \frac{1}{120}\left( \left|\frac{kQ_{yz}^m}{c}\right|^2 + \left|\frac{kQ_{zy}^m}{c}\right|^2 \right) \right] \quad (3)$$

EQ + MQ: $$C_{sca}^{total} = C_{sca}^{Q^e} + C_{sca}^{Q^m} = \frac{k^4}{6\pi\varepsilon_0^2 |\mathbf{E}_{inc}|^2}\left[ \frac{1}{120}\left( |kQ_{xz}^e|^2 + |kQ_{zx}^e|^2 + \left|\frac{kQ_{yz}^m}{c}\right|^2 + \left|\frac{kQ_{zy}^m}{c}\right|^2 \right) \right] \quad (4)$$

We proceed in a similar way with the formula (9) in main part of manuscript for the scattering cross-section of multipoles ($n_z$ = -1, $E_x^{bwsc} = 0$, because forward scattering only).

ED + MD: $$E_x^{bwsc} = \frac{k^2}{4\pi\varepsilon_0}\frac{e^{ikr}}{r}\left\{ p_x - \frac{1}{c} m_y \right\} = 0 \quad (5)$$

ED + EQ: $$E_x^{bwsc} = \frac{k^2}{4\pi\varepsilon_0}\frac{e^{ikr}}{r}\left\{ p_x + \frac{ik}{6} Q_{xz}^e \right\} = 0 \quad (6)$$

MD + MQ: $$E_x^{bwsc} = \frac{k^2}{4\pi\varepsilon_0}\frac{e^{ikr}}{r}\left\{ -\frac{1}{c} m_y - \frac{ik}{6c} Q_{yz}^m \right\} = 0 \quad (7)$$

EQ + MQ: $$E_x^{bwsc} = \frac{k^2}{4\pi\varepsilon_0}\frac{e^{ikr}}{r}\left\{ \frac{ik}{6} Q_{xz}^e - \frac{ik}{6c} Q_{yz}^m \right\} = 0 \quad (8)$$

Now we can take the multipoles moments from the formulas and substitute them into the formulas for the scattering cross-section to obtain conditions in the absence of backscattering.

ED + MD: $$C_{sca}^{total} = C_{sca}^{p} + C_{sca}^{m} = \frac{k^4}{6\pi\varepsilon_0^2 |\mathbf{E}_{inc}|^2}\left[ \left|\frac{m_y}{c}\right|^2 + \left|\frac{m_y}{c}\right|^2 \right] \quad (9)$$

ED + EQ: $$C_{sca}^{total} = C_{sca}^{p} + C_{sca}^{Q^e} = \frac{k^4}{6\pi\varepsilon_0^2 |\mathbf{E}_{inc}|^2}\left[ \frac{1}{36}|-ikQ_{xz}^e|^2 + \frac{1}{60}\left( |kQ_{xz}^e|^2 \right) \right] \quad (10)$$

MD + MQ: $$C_{sca}^{total} = C_{sca}^{m} + C_{sca}^{Q^m} = \frac{k^4}{6\pi\varepsilon_0^2 |\mathbf{E}_{inc}|^2}\left[ \frac{1}{36}\left|-\frac{ikQ_{yz}^m}{c}\right|^2 + \frac{1}{60}\left( \left|\frac{kQ_{yz}^m}{c}\right|^2 \right) \right] \quad (11)$$

EQ + MQ: $$C_{sca}^{total} = C_{sca}^{Q^e} + C_{sca}^{Q^m} = \frac{k^4}{6\pi\varepsilon_0^2 \left|\mathbf{E}_{inc}\right|^2}\left[\frac{1}{60}\left(\left|\frac{kQ_{yz}^m}{c}\right|^2 + \left|\frac{kQ_{yz}^m}{c}\right|^2\right)\right] \quad (12)$$

This shows how the scattering cross-sections of different multipoles must be related to each other to fulfill the conditions for the absence of backscattering. Similar calculations can be used to obtain the same values for the absence of forward scattering.

ED + MD: $$\frac{C_{sca}^{ED}}{C_{sca}^{MD}} = 1 \quad (13)$$

ED + EQ: $$\frac{C_{sca}^{ED}}{C_{sca}^{EQ}} = 1.67 \quad (14)$$

MD + MQ: $$\frac{C_{sca}^{MD}}{C_{sca}^{MQ}} = 1.67 \quad (15)$$

EQ + MQ: $$\frac{C_{sca}^{EQ}}{C_{sca}^{MQ}} = 1 \quad (16)$$

Following the same logic as for the Generalized Kerker effect, we obtain the conditions for the Transverse Kerker effect.

ED + MQ: $$E_x^{bwsc} = E_x^{fwsc} = \frac{k^2}{4\pi\varepsilon_0}\frac{e^{ikr}}{r}\left\{p_x - \frac{ik}{6c}Q_{yz}^m\right\} = 0 \quad (17)$$

MD + EQ: $$\begin{cases} E_x^{bwsc} = \dfrac{k^2}{4\pi\varepsilon_0}\dfrac{e^{ikr}}{r}\left\{-\dfrac{1}{c}m_y + \dfrac{ik}{6}Q_{xz}^e\right\} = 0 \\ E_x^{fwsc} = \dfrac{k^2}{4\pi\varepsilon_0}\dfrac{e^{ikr}}{r}\left\{\dfrac{1}{c}m_y - \dfrac{ik}{6}Q_{xz}^e\right\} = 0 \end{cases} \quad (18)$$

Then

ED + MQ: $$C_{sca}^{total} = C_{sca}^{p} + C_{sca}^{Q^m} = \frac{k^4}{6\pi\varepsilon_0^2 \left|\mathbf{E}_{inc}\right|^2}\left[\frac{1}{36}\left|\frac{ikQ_{yz}^m}{c}\right|^2 + \frac{1}{60}\left(\left|\frac{kQ_{yz}^m}{c}\right|^2\right)\right] \quad (19)$$

MD + EQ: $$C_{sca}^{total} = C_{sca}^{m} + C_{sca}^{Q^e} = \frac{k^4}{6\pi\varepsilon_0^2 \left|\mathbf{E}_{inc}\right|^2}\left[\frac{1}{36}\left|-ikQ_{xz}^e\right|^2 + \frac{1}{60}\left(\left|kQ_{xz}^e\right|^2\right)\right] \quad (20)$$

Whence it follows that:

ED + MQ: $$\frac{C_{sca}^{ED}}{C_{sca}^{MQ}} = 1.67 \quad (21)$$

MD + EQ: $$\frac{C_{sca}^{MD}}{C_{sca}^{EQ}} = 1.67 \quad (22)$$

## S3. Specific parameters of the considered effects for truncated silicon nanocones.

These data will make it easy to get any listed effect at the desired wavelength using the following simple relationships, provided that the refractive indices at these wavelengths are approximately equal:

$$\begin{cases} \dfrac{R_{top}}{R_{bottom}} n \approx \dfrac{R_{top}^*}{R_{bottom}^*} n^* \approx const \\ \dfrac{H}{\lambda} n \approx \dfrac{H^*}{\lambda^*} n^* \approx const \\ \dfrac{R}{\lambda} n \approx \dfrac{R^*}{\lambda^*} n^* \approx const \end{cases} \quad (23)$$

, where the parameters without an asterisk are the parameters below and the parameters with an asterisk are the parameters to be retrieved.

| | **Generalized** | **Transverse** |
|---|---|---|
| **ED + MD** | H = 140 nm<br>$R_{top}$ = 120 nm<br>$R_{bottom}$ = 90 nm<br>λ = 779 nm | - |
| **ED + EQ** | H = 160 nm<br>$R_{top}$ = 100 nm<br>$R_{bottom}$ = 250 nm<br>λ = 675 nm | - |
| **ED + MQ** | - | H = 420 nm<br>$R_{top}$ = 130 nm<br>$R_{bottom}$ = 110 nm<br>λ = 779 nm |
| **MD + EQ** | - | H = 300 nm |
| | | $R_{top}$ = 40 nm<br>$R_{bottom}$ = 120 nm<br>$\lambda$ = 617 nm |
| **MD + MQ** | H = 100 nm<br>$R_{top}$ = 190 nm<br>$R_{bottom}$ = 140 nm<br>$\lambda$ = 767 nm | - |
| **EQ + MQ** | H = 320 nm<br>$R_{top}$ = 360 nm<br>$R_{bottom}$ = 180 nm<br>$\lambda$ = 815 nm | - |
| **ED + MD + EQ + MQ** | H = 220 nm<br>$R_{top}$ = 50 nm<br>$R_{bottom}$ = 330 nm<br>$\lambda$ = 625 nm | H = 520 nm<br>$R_{top}$ = 55 nm<br>$R_{bottom}$ = 195 nm<br>$\lambda$ = 900 nm |

**Supplementary Table S1.** Specific parameters of nano-scatterers for obtaining Kerker effects for truncated silicon nanocones.

**Hybrid Anapole regime**

H = 200 nm, $R_{top}$ = 40 nm, $R_{bottom}$ = 340 nm, $\lambda$ = 625 nm (truncated cone)

H = 369 nm, R = 127 nm, $\lambda$ = 748 nm (cylinder)

**Superscattering**

P1: H = 415 nm, $R_{top}$ = 216 nm, $R_{bottom}$ = 180 nm, $\lambda$ = 760 nm

P2: H = 415 nm, $R_{top}$ = 144 nm, $R_{bottom}$ = 180 nm, $\lambda$ = 673 nm

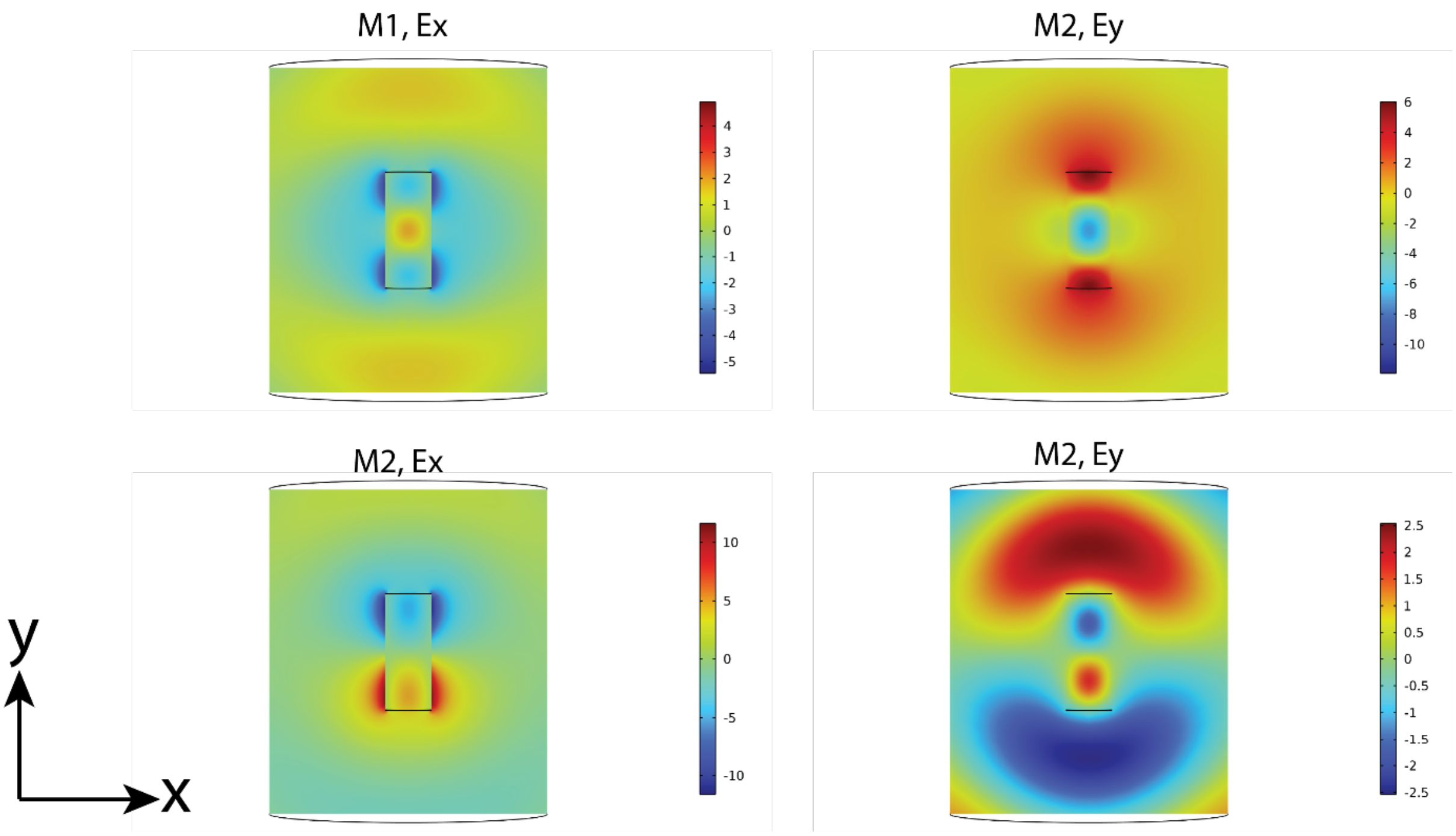

**Supplementary Figure S2.** QNMs M1 and M2 field distributions for the cylinder presented in figure 10 in the main text. The field distributions show M1 is an even-parity mode while M2 is an odd-parity mode.